\documentclass[letterpaper,twocolumn,10pt]{article}
\usepackage{hyperref}
\usepackage{usenix}

\usepackage{xcolor,colortbl}
\usepackage[english]{babel}
\usepackage{blindtext}
\usepackage{multicol}

\usepackage{algorithm}
\usepackage{algorithmicx}
\usepackage{algpseudocode}
\usepackage{amsmath}

\usepackage{tikz}
\usepackage{amsmath}
\usepackage{xspace}
\usepackage{booktabs}
\usepackage{subcaption}
\usepackage{multirow}

\usepackage{tabularx}

\usepackage[subtle]{savetrees}
\usepackage[small, compact]{titlesec}

\usepackage{inconsolata}

\usepackage{siunitx}
\usepackage{pifont}

\def\Snospace~{\S{}}

\newcommand{\paraspace}{\vspace{0.05in}}
\newcommand{\parab}[1]{\paraspace\noindent{\bf #1} }

\newcommand{\us}{\textmu s\xspace}

\newcommand{\autorefsuffix}[2]{\hyperref[#1]{\autoref*{#1}#2}}

\newcommand{\sys}{mRPC\xspace}
\newcommand{\sysservice}{\sys service\xspace}

\newcommand{\datapath}{datapath\xspace}
\newcommand{\datapaths}{datapaths\xspace}

\usepackage{enumitem}

\begin{document}
\pagenumbering{gobble}
\title{Remote Procedure Call as a Managed System Service}

\author{
\rm{Jingrong Chen$^{\text{1}, *}$\enskip
    Yongji Wu$^{\text{1}, *}$ \enskip
    Shihan Lin$^{\text{1}}$ \enskip
    Yechen Xu$^{\text{3}}$ \enskip
    Xinhao Kong$^{\text{1}}$ \enskip} \\
\rm{Thomas Anderson$^{\text{2}}$ \enskip
    Matthew Lentz$^{\text{1}}$ \enskip
    Xiaowei Yang$^{\text{1}}$ \enskip
    Danyang Zhuo$^{\text{1}}$ \enskip}\\ \\
{$^{\text{1}}$Duke University\enskip$^{\text{2}}$University of Washington\enskip$^{\text{3}}$Shanghai Jiao Tong University}
}

\maketitle
\begin{abstract}

Remote Procedure Call (RPC) is a widely used abstraction for cloud computing. 
The programmer specifies type information for each remote procedure, and a compiler generates stub code linked into each application 
to marshal and unmarshal arguments into message buffers. Increasingly, however,
application and service operations teams need a high degree of visibility and control over the flow of RPCs between services, leading
many installations to use sidecars or service mesh proxies for manageability and policy flexibility. These sidecars typically involve
inspection and modification of RPC data that the stub compiler had just carefully assembled, adding
needless overhead. Further, upgrading diverse application RPC stubs to use advanced hardware capabilities such as RDMA or DPDK is a long and involved process, and often incompatible with sidecar policy control.

In this paper, we propose, implement, and evaluate a novel approach, where RPC marshalling and policy enforcement are done
as a system service rather than as a library linked into each application. Applications specify type information
to the RPC system as before, while the RPC service executes policy engines and arbitrates resource use, and then marshals data
customized to the underlying network hardware capabilities. Our system, \sys, also supports live upgrades so that both policy and marshalling code can be updated transparently to application code.  
Compared with using a sidecar,
\sys speeds up a standard microservice benchmark, DeathStarBench, by up to 2.5$\times$
while having a higher level of policy flexibility and availability.
\end{abstract}
{\let\thefootnote\relax\footnote{{$^*$Jingrong Chen and Yongji Wu contributed equally.}}}
\section{Introduction}

Remote Procedure Call (RPC) is a fundamental building block of distributed systems in modern datacenters. 
RPC allows developers to build networked applications using a simple and familiar programming model~\cite{birrell1984rpc}, 
supported by several popular libraries such as gRPC~\cite{grpc}, Thrift~\cite{slee2007thrift}, and eRPC~\cite{kalia2019erpc}.
The RPC model has been widely adopted in distributed data stores~\cite{kalia2016fasst, singhvi2021cliquemap, etcd}, network file systems~\cite{sandberg1986sunnfs, gluster}, consensus protocols~\cite{ongaro2014raft}, data-analytic frameworks~\cite{dean2004mapreduce, shvachko2010hdfs, zaharia2012rdd, carbone2015flink, toshniwal2014storm, arrow, low2012graphlab, gonzalez2012powergraph}, cluster schedulers and orchestrators~\cite{kubernetes, hindman2011mesos}, and machine learning systems~\cite{paszke2019pytorch, abadi2016tensorflow, philipp2018ray}. 
Google found that roughly 10\% of its datacenter CPU cycles are spent just executing gRPC library code~\cite{kanev2015datacentertax}.
Because of its importance, improving RPC performance has long been a major topic of research~\cite{birrell1984rpc, schroeder1989firefly, bershad1989lrpc, bershad1991urpc, vadhat2020keynote, stuedi2014darpc, tsai2017lite, kalia2019erpc, monga2021flock, li2021hatrpc, cho2020overload}. 

Recently, application and network operations teams have found a need for rapid and flexible visibility and control 
over the flow of RPCs in datacenters. 
This includes monitoring and control of the performance of specific types of RPCs~\cite{nines}, 
prioritization and rate limiting to meet application-specific performance and availability goals,
dynamic insertion of advanced diagnostics to track user requests across a network of microservices~\cite{xtrace}, 
and application-specific load balancing to improve cache effectiveness~\cite{cachelib}, to name a few. 

\begin{figure}
\centering
\includegraphics[width=0.98\linewidth]{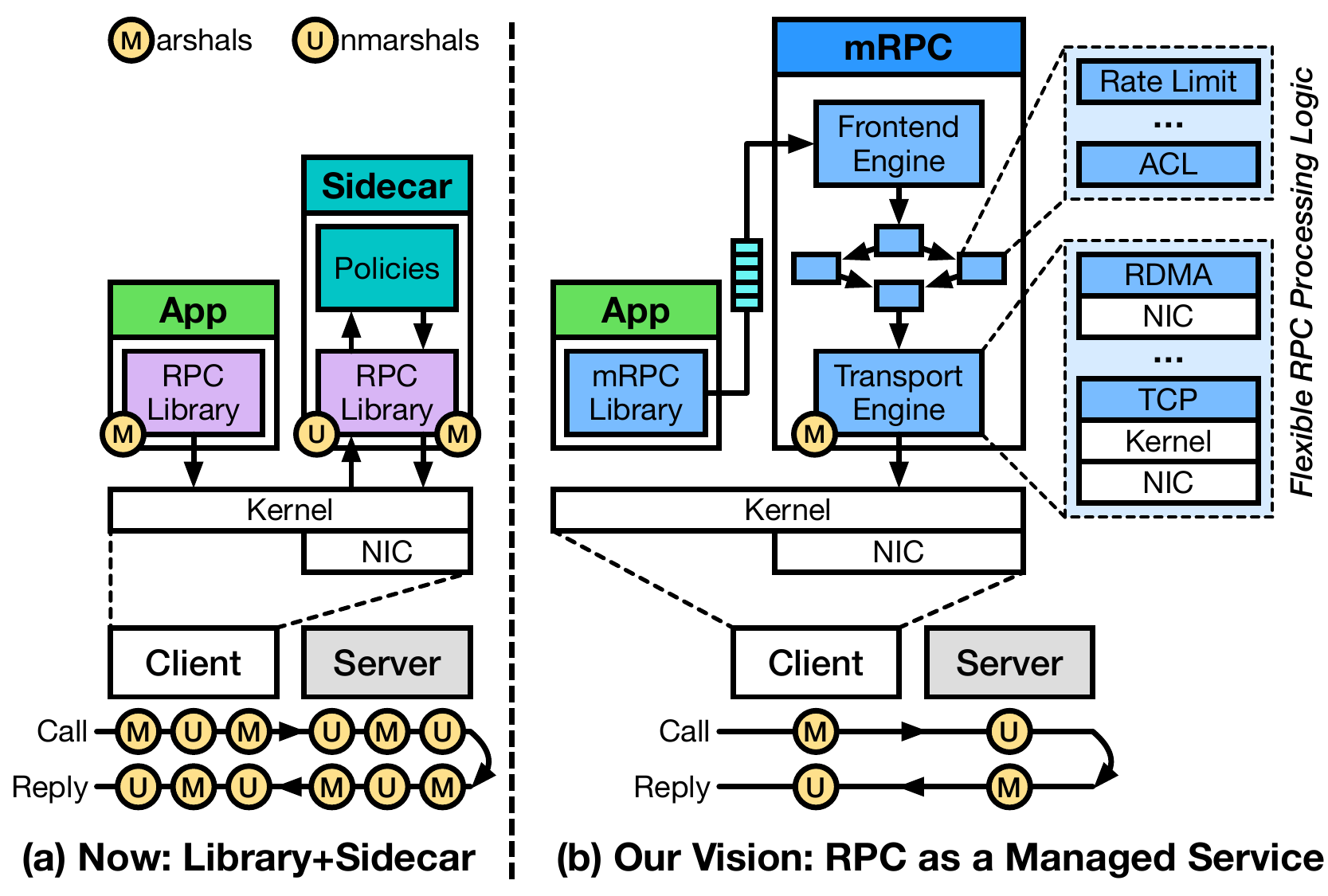}
\vspace{-3mm}
\caption{Architectural comparison between current (RPC library + sidecar) and our proposed (RPC as a managed service) approaches.}
\vspace{-4mm}
\label{fig:comparison}
\end{figure}

The typical architecture is to enforce policies in a sidecar---a separate process that mediates the network traffic of the application RPC library (\autorefsuffix{fig:comparison}{a}). This is often referred to as a
service mesh. A number of commercial products
have been developed to meet the need for sidecar RPC proxies, such as 
Envoy~\cite{envoy}, Istio~\cite{istio}, HAProxy~\cite{haproxy}, Linkerd~\cite{linkerd}, Nginx~\cite{nginx},
and Consul~\cite{consul}.
Although some policies could theoretically be supported by a feature-rich RPC runtime linked in with each application, 
that can slow deployment---Facebook recently reported that it can take 
{\em months} to fully roll out changes to one of its application communication libraries~\cite{owl}.
One use case that requires rapid deployment is to respond to a new application security threat, or to diagnose and fix
a critical user-visible failure.  Finally, many policies are mandatory rather than discretionary---the network 
operations team may not be able to trust the library code linked into an application.
Example mandatory security policies include access control, authentication/encryption~\cite{consul}, and prevention of
known exploits in widely used network protocols such as RDMA~\cite{redmark}.

Although using a sidecar for policy management is
functional and secure, 
it is also inefficient.
The application RPC library marshals 
RPC parameters at runtime into a buffer according to the type information provided by the programmer. This buffer is sent through the operating system network stack and then forwarded back up to the sidecar, 
which typically needs to parse and unwrap the network, virtualization, and RPC headers,
often looking inside 
the packet payload to correctly enforce the desired policy. It then re-marshals the data for transport.
Direct application-level access to network 
hardware such as RDMA or DPDK offers high performance but precludes sidecar policy control.  
Similarly, network interface cards are increasingly sophisticated, but it is hard for applications or sidecars
to take advantage of those new features, because marshalling is done too high up in the network stack.
Any change to the marshalling code requires recompiling and rebooting each application and/or the sidecar,
hurting end-to-end availability.
In short, existing solutions can provide good performance, or flexible and enforceable policy control, but not both.

In this paper, we propose a new approach, called RPC as a managed service, to address these limitations.
Instead of separating marshalling and policy enforcement across different domains, we combine 
them into a single privilege and trusted
system service (\autorefsuffix{fig:comparison}{b}) so that marshalling is done \textbf{after} policy processing. 
In our prototype, mRPC for managed RPC, the privileged RPC service runs at user level
communicating with the application through shared memory regions~\cite{bershad1991urpc,multikernel,marty2019snap}.  
However, mRPC could also be integrated directly into the operating system kernel 
with a dynamically replaceable kernel module~\cite{miller2021bento}. 

Our goals are to be fast, support flexible policies, and provide high availability for applications.
To achieve this, we need to address several challenges. First, we need to decouple marshalling from the application RPC library.
Second, we need to design a new policy enforcement mechanism to process RPCs efficiently and securely, without incurring additional marshalling overheads. Third, we need to provide a way for operators to specify/change policies and even change the underlying transport implementation without disrupting running applications. 

We implement \sys, the first RPC framework that follows the RPC as a managed service approach.
Our results show that
\sys speeds up DeathStarBench~\cite{gan2019deathstarbench} by up to 2.5$\times$, in terms of mean latency,
compared with combining state-of-art RPC libraries and sidecars, i.e., gRPC and Envoy, using
the same transport mechanism. Larger performance gains are possible by fully exploiting network hardware capabilities from within
the service. 
In addition, \sys allows for live upgrades of its components while incurring negligible downtime for applications. Applications do not need to be re-compiled or rebooted to change policies or marshalling code. \sys has three important limitations. 
First, data structures passed as RPC arguments must be allocated on a special shared-memory heap.  Second, while we use
a language-independent protocol for specifying RPC type signatures, our prototype implementation currently only works with
applications written in Rust. Finally, our stub generator is not as fully featured as gRPC. 

In this paper, we make the following contributions:
\begin{itemize}
    \item A novel RPC architecture that decouples marshalling/unmarshalling from RPC libraries to a centralized system service.
    \item An RPC mechanism that applies network policies and observability features with both security and low performance overhead, i.e., with minimal data movement and no redundant (un)marshalling. The mechanism supports live upgrade of 
    RPC bindings, policies, transport, and marshalling without disrupting running applications.
\vspace{-1.5ex}    \item A prototype implementation of \sys along with an evaluation on both synthetic workloads and real applications.
\end{itemize}
\section{Background}
\label{sec:background}

In this section, we discuss the current RPC library architecture. We then discuss the emerging need for manageability and how manageability is implemented with existing RPC libraries.

\subsection{Remote Procedure Call}

To use RPC, a developer defines the relevant service interfaces and message
types in a schema file (e.g., gRPC \texttt{.proto} file).
A protocol compiler will translate the schema into program stubs that are
directly linked with the client and server applications.
To issue an RPC at runtime, the application simply calls the corresponding
function provided by the stub; the stub is responsible for marshalling the
request arguments and interacting with the transport layer (e.g., TCP/IP sockets or
RDMA verbs).
The transport layer delivers the packets to the remote server, where the
stub unmarshals the arguments and dispatches the RPC request to a thread (eventually replying back to the client).
We refer to this approach as \emph{RPC-as-a-library}, since all RPC
functionality is included in user-space libraries that are linked with
each application.
Even though the first RPC implementation~\cite{birrell1984rpc} dates back to
the 1980s, modern RPC frameworks (e.g., gRPC~\cite{grpc},
eRPC~\cite{kalia2019erpc}, Thrift~\cite{slee2007thrift}) still follow this
same approach.

A key design goal for RPC frameworks is efficiency.
Google and Facebook have built their own efficient RPC
frameworks, gRPC and Apache Thrift. Although
primarily focused on portability and interoperability, gRPC includes many efficiency-related
features, such as supporting binary payloads. Academic researchers have
studied various ways to improve RPC efficiency, including
optimizing the network stack~\cite{kaufmann2019tas,zhang2021demikernel,ousterhout2019shenango}, software hardware
co-design~\cite{kalia2019erpc, kalia2016fasst}, and overload
control~\cite{cho2020overload}. 

As network link speeds continue to scale up~\cite{jupiterevolving}, RPC overheads are likely to become
even more salient in the future.  This has led some researchers to advocate for direct application access
to network hardware~\cite{peter2014arrakis,belay2014ix,kalia2019erpc,zhang2021demikernel}, e.g., with RDMA or DPDK.  
Although low overhead, kernel bypass is largely incompatible with the need for flexible and enforceable layer 7 policy control,
as we discuss next. In practice, multiple security weaknesses in RDMA hardware have led most cloud vendors to opt against 
providing direct access to RDMA by untrusted applications~\cite{tsai2017lite,marty2019snap,redmark,kong2023husky,kong2022collie, zhang2022justitia}.

\begin{figure*}[t]
\centering
\includegraphics[width=.95\linewidth]{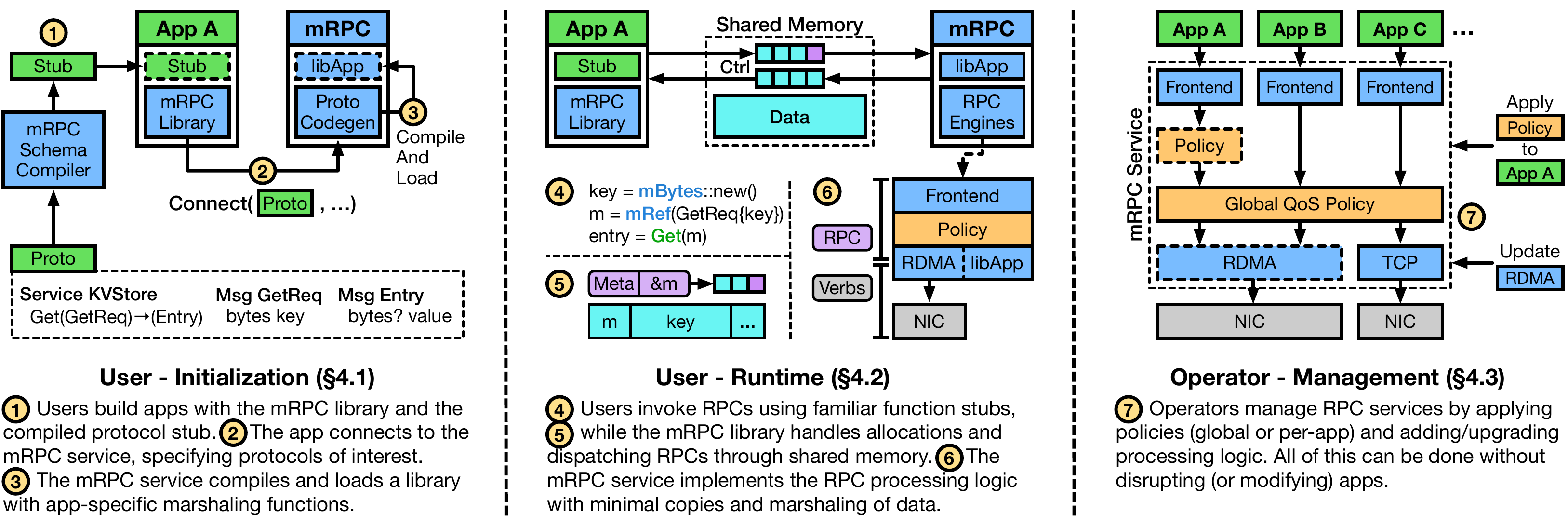}
\vspace{-2mm}
\caption{Overview of the mRPC workflow from the perspective of the users
  (and their applications) as well as infrastructure operators.}
\vspace{-4mm}
\label{fig:overview}
\end{figure*}

\subsection{The Need for Manageability}

As RPC-based distributed applications scale to large, complex deployment
scenarios, there is an increasing need for improved manageability of RPC
traffic. We classify management needs into three categories: \textbf{1) Observability:} Provide detailed telemetry, which enables developers to diagnose and optimize application performance. \textbf{2) Policy Enforcement:} Allow operators to apply custom policies to RPC applications and services (e.g., access control, rate limits, encryption). \textbf{3) Upgradability:} Support software upgrades (e.g., bug fixes and new features) while minimizing downtime to applications.

One natural question to ask is: \textit{is it possible to add these
properties without changing existing RPC libraries?} For
observability and policy enforcement, the state-of-the-art solution is to
use a sidecar (e.g., Envoy~\cite{envoy} or Linkerd~\cite{linkerd}).
A sidecar is a standalone process that intercepts every packet an
application sends, reconstructing the application-level data (i.e., RPC),
and applying policies or enabling observability.
However, using a sidecar introduces substantial performance overhead,
due to redundant RPC (un)marshalling.
This RPC (un)marshalling, for example, in gRPC+Envoy, including HTTP framing and protobuf encoding,
accounts for 62-73\% overhead in the end-to-end latency~\cite{zhu2022dissecting}. In our evaluation (\autoref{sec:eval}), using a sidecar increases the 99th percentile RPC latency by 180\% and decreases the bandwidth by 44\%.
\autorefsuffix{fig:comparison}{a}
shows the (un)marshalling steps invoked as an RPC traverses from
a client to a server and back. Using a sidecar triples the
number of (un)marshalling steps (from 4 to 12). In addition, the sidecar
approach is largely incompatible with the emerging trend of efficient
application-level access to network hardware. Using
sidecars means data buffers have to be copied between the application and
sidecars, reducing the benefits of having zero-copy kernel-bypass access to the
network.

Finally, using sidecars with application RPC libraries does not completely solve the 
upgradability issue. While policy can often be changed dynamically (depending on the feature set
of the sidecar implementation), marshalling and transport code is harder to change. 
To fix a bug in the underlying RPC library, or merely to upgrade the code to take advantage
of new hardware features, we need to recompile the entire application (and sidecar) with the patched RPC library and reboot.
gRPC has a monthly or two-month release cycle for bug fixes and new
features~\cite{grpc-release-schedule}. Any scheduled downtime has to be
communicated explicitly to the users of the application or has to be masked
using replication; either approach can lead to complex application
life-cycle management issues.

\textit{We do not see much hope in continuing to optimize this RPC library and sidecar approach for two reasons.} First, a strong coupling exists between a traditional RPC library and each application. This makes upgrading the RPC library without stopping the application difficult, if not impossible. Second, there is only weak or no coupling between an RPC library and a sidecar. This prevents the RPC library and the sidecar from cross-layer optimization.

Instead, we argue for an alternative architecture in which RPC is provided \textit{as a managed service}.
By decoupling RPC logic, e.g., (un)marshalling, transport interface, from the application, the service can simultaneously provide high performance, policy flexibility, and zero-downtime upgrades.
\section{Overview}
\label{sec:overview}

\newcommand*\circled[1]{\tikz[baseline=(char.base)]{
  \node[shape=circle,draw,inner sep=1pt,fill=yellow,thick] (char)
  {\footnotesize\textbf #1};}}

Our system, mRPC, realizes the \emph{RPC-as-a-managed-service} abstraction
while maintaining similar end-to-end semantics as traditional
RPC libraries (e.g., gRPC, Thrift).
The goals for mRPC are to be fast, support flexible policy enforcement, and provide
high availability for applications.

\autoref{fig:overview} shows a high-level overview of the mRPC architecture
and workflow, breaking it down into three major phases: initialization,
runtime, and management.
The mRPC service runs as a non-root, user-space process with
access to the necessary network devices and a shared-memory region for each application.
In each of the phases, we focus on the view of a single machine that is
running both the RPC client application and the mRPC service.
The RPC server may also run alongside an mRPC service. In this case, mRPC-specific marshalling can be used.
However, we also support flexible marshalling to enable mRPC applications to interact with external peers using well-known formats (e.g., gRPC).
In our evaluation, we focus on cases where both the client and server employ mRPC.

The initialization phase extends from building the application to how the
application binds to a specific RPC interface.
\circled{1} Similar to gRPC, users define a protocol schema. The mRPC
schema compiler uses this to generate stub code to include in their application.
We illustrate this using a key-value storage service with a single
\texttt{Get} function.
\circled{2} When the application is deployed, it connects with the mRPC
service running on the same machine and specifies the protocol(s) of
interest, which are maintained by the generated stub.
\circled{3} The mRPC service also uses the protocol schema to generate, compile, and dynamically 
load a protocol-specific library containing the marshalling and unmarshalling code for
that application's schemas\footnote{Note that such libraries may be
prefetched and/or cached to optimize the startup time.}.
This \emph{dynamic binding} is a key enabler for mRPC to act as a
long-running service, handling arbitrary applications (and their RPC
schemas).
\footnote{The dashed box of "Stub" and "libApp" means they are generated code.}

At this point, we enter the runtime phase in which the application begins to
invoke RPCs.
Our approach uses \textit{shared memory} between the application
and mRPC, containing both control queues as well as a data buffer.
\circled{4} The application protocol stub produced by the mRPC protocol compiler can be called
like a traditional RPC interface, with the exception that data structures passed as arguments or
as return values must be allocated on a special heap in the shared data buffer. As an example, we show an
excerpt of Rust-like pseudocode for invoking the \texttt{Get} function.
\circled{5} Internally, the stub and mRPC library manage RPC calls and
replies in the control queues along with allocations and deallocations in the data buffer.
\circled{6} The mRPC service operates over the RPCs through modular
\emph{engines} that are composed to implement the per-application \emph{\datapaths} (i.e., sequence of RPC processing logic); each engine is responsible for one type of task (e.g., application interface, rate limiting, transport interface).
Engines do not contain execution contexts, but are rather scheduled by \emph{runtimes} in \sys that correspond to kernel-level threads; during their execution, engines read from input queues, perform work, and enqueue outputs.
External-facing engines (i.e., frontend, transport) use asynchronous control queues, while all other engines are executed synchronously by a runtime.
Application control queues are contained in shared memory with the \sysservice.

This architecture, along with dynamic binding, enables mRPC to \emph{operate
over RPCs rather than packets}, avoiding the high overhead
of traditional sidecar-based approaches.
Additionally, the modular design of \sys's processing logic enables mRPC to take
advantage of fast network hardware (e.g., RDMA and smartNICs) in a manner that
is transparent to the application.
A key challenge, which we will address in \autoref{sec:manage-with-efficiency}, is how to
securely enforce operator policies over RPCs in shared memory while
minimizing data copies.

Finally, mRPC aims to improve the manageability of RPCs by infrastructure
operators.
Here, we zoom out to focus on the processing logic across all
applications served by an mRPC service.
\circled{7} Operators may wish to apply a number of different policies to
RPCs made by applications, whether on an individual basis (e.g., rate
limiting, access control) or globally across applications (e.g., QoS).
mRPC allows operators to add, remove, update, or reconfigure policies at
runtime.
This flexibility extends beyond policies to
include those responsible for interacting with the network hardware.
A key challenge, which we will address in \autoref{sec:upgrade}, is in supporting
the \emph{live upgrade} of mRPC engines without interrupting running
applications (and while managing engines sharing memory queues).

\section{Design}
\label{sec:design}

In this section, we describe how \sys provides dynamic binding, efficient policy and observability support, live upgrade, and security. 

\subsection{Dynamic RPC Binding}
\label{sec:binding}

Applications have different RPC schemas, which ultimately decide how an RPC is marshalled.
In the traditional RPC-as-a-library approach, a protocol compiler generates the marshalling code, which is linked into the application.
In our design, the \sysservice is responsible for marshalling, which means that the application-specific marshalling code needs to be decoupled from an RPC library and run inside the \sysservice itself.
Failing to ensure this separation would allow arbitrary code execution by a malicious user.

Applications directly submit the RPC schema (and not marshalling code) to the \sysservice. The \sysservice generates the corresponding marshalling code, then compiles and dynamically loads the library. Thus, we rely on our \sysservice code generator to produce the correct marshalling code for \textit{any} user-provided RPC schema. For the initial handshake between an RPC client and an RPC server, the two mRPC services check that the provided RPC schemas match, and if not, the client's connection is rejected.

There are three remaining questions. First, \textbf{what are the responsibilities of the in-application user stub and mRPC library?} In \sys, applications rely on user stubs to implement the abstraction as specified in their RPC schema. This means we still need to generate the glue code to maintain the traditional application programming interface. Our solution is to provide a separate protocol schema compiler, which is untrusted and run by application developers, to generate the user stub code that does not involve marshalling and transport.
The application RPC stub (with the help of the mRPC library) creates a message buffer that contains the metadata of the RPC, with typed pointers to the RPC arguments, on the shared memory heap. The message is placed on a shared memory queue, which will be processed by the \sysservice. The receiving side works in a similar way.

Second, \textbf{does this approach increase RPC connect/bind time?} Implemented naively, this design will increase the RPC connect/bind time because the \sysservice has to compile the RPC schema and load the resulting marshalling library when an RPC client first connects to a corresponding server (or equivalently when an RPC server binds to the service). However, this latency is not fundamental to our design, and we can mitigate it in the following way. The \sysservice accepts RPC schemas before booting an application, as a form of prefetching.  Given a schema, it compiles and caches the marshalling code. At the time of RPC connect/bind, the \sysservice simply performs a cache lookup based on the hash of the RPC schema. If it exists within the cache, the \sysservice will load the associated library; otherwise, the \sysservice will invoke the compiler to generate (and subsequently cache) the library. This reduces the connect/bind time from several seconds to several milliseconds.

Third, \textbf{when new applications arrive, do existing applications face downtime?} The multi-threaded \sysservice is a single process that serves many RPC applications;
however, the marshalling engines for different RPC applications are not shared. 
They are in different memory addresses and can be (un)loaded independently.
We will describe in \autoref{sec:upgrade} how to load/unload engines without disrupting running applications.

\subsection{Efficient RPC Policy Enforcement and Observability}
\label{sec:manage-with-efficiency}

We have one key idea to allow efficient RPC policy enforcement and observability: senders should marshal once (as late as possible), while receivers should unmarshal once (as early as possible).
On the sender side, we want to support policy enforcement and observability directly over RPCs from the application, and then marshal the RPC into packets. The receiver side is similar: packets should be unmarshalled into RPCs, applying policy and observability operations, and then delivered directly to the application. Compared to the traditional RPC-as-a-library approach with sidecars, this eliminates the redundant (un)marshalling steps (see \autoref{fig:comparison}).

\parab{Data: DMA-capable shared memory heaps.}
Our design is centered around a dedicated shared memory heap between each application and the \sysservice. (Note that this heap is not shared across applications.) Applications directly create data structures, which may be used in RPC arguments, in a shared memory heap with the help of the \sys library. Each application has a separate shared memory region, which provides isolation between (potentially mutually distrusting) applications. The \sys library also includes a standard slab allocator for managing object allocation on this shared memory. If there is insufficient space within the shared memory, the slab allocator will request additional shared memory from the \sysservice and then map it into the application's address space. The \sysservice has access to the shared memory heap, allowing it to execute RPC processing logic over the application's RPCs, but also maintains a private memory heap for necessary copies.

\begin{figure}[t]
  \centering
  \includegraphics[width=0.98\linewidth]{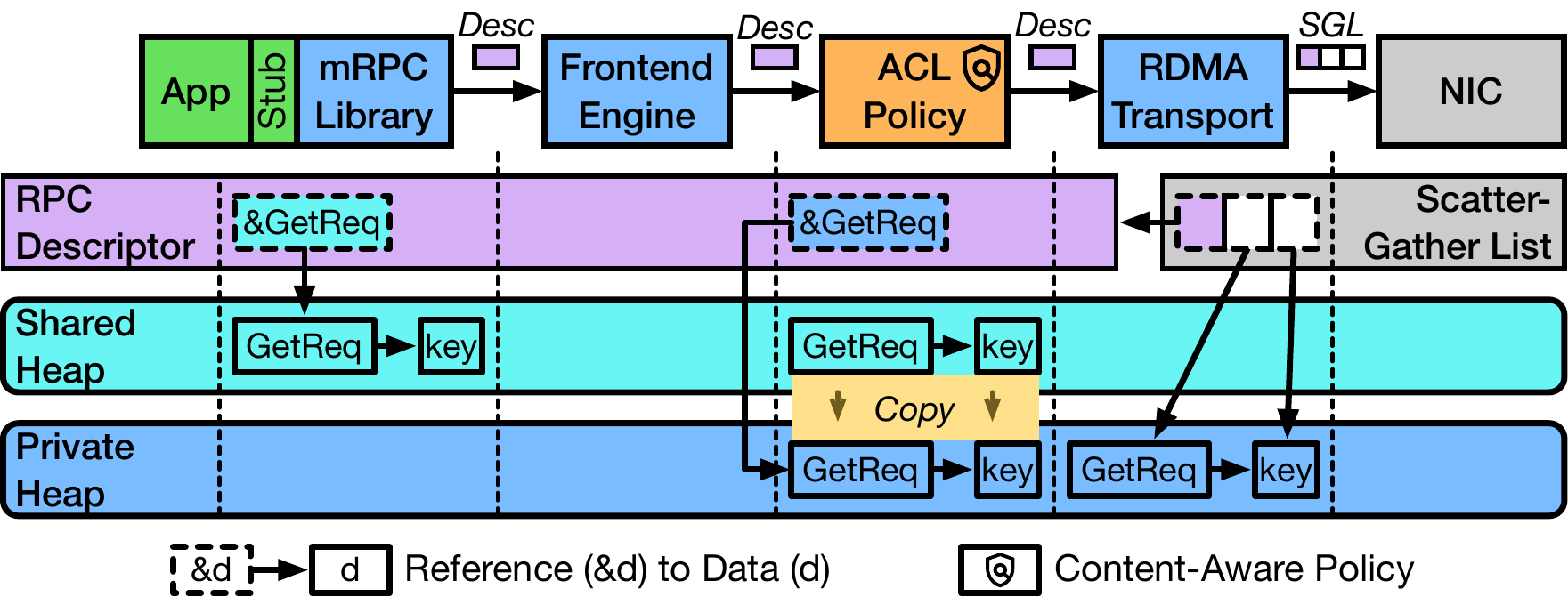}
\vspace{-2mm}
  \caption{Overview of memory management in mRPC. Shows an example for the
  \texttt{Get} RPC that includes a content-aware ACL policy.} 
\vspace{-4mm}
  \label{fig:mrpc-shared-heap}
\end{figure}

\autoref{fig:mrpc-shared-heap} shows an example workflow that includes access control for a key-value store service.
Having the data structures directly in the shared memory allows an application to provide pointers to data, rather than the data itself, when submitting RPCs to the \sysservice. We call the message sent from an application to the \sysservice an \emph{RPC descriptor}. If there are multiple RPC arguments, the RPC descriptor points to an array of pointers (each pointing to a different argument on the heap).

Let us say we have an ACL policy that rejects an RPC if the key matches a certain string. The \sysservice first copies the argument (i.e., \texttt{key}), as well as all parental data structures (i.e., \texttt{GetReq}), onto its private heap. This is to prevent time-of-use-to-time-of-check (TOCTOU) attacks. Since applications have access to DMA-capable shared memory at all times, an application could modify the content in the memory while the \sysservice is enforcing policies. Copying arguments is a standard mitigation technique, similar to how OS kernels prevents TOCTOU attacks by copying system call arguments from user- to kernel-space. This copying only needs to happen if the policy behavior is based on the content of the RPC. We demonstrate in \autoref{sec:policy} that even with such copying, mRPC's overhead for an ACL policy is much lower than gRPC + Envoy.  The RPC descriptor is modified so that the pointer to the copied argument now points to the private heap.
On the receiver side, the TOCTOU attack is not relevant, but we need to take care not to place RPCs directly in shared memory. 
If there is a receive-side policy that depends on RPC argument values, the \sysservice first receives the RPC data into a private heap; it copies the RPC data into the shared heap after policy processing.
This prevents the application from reading RPC data that should have been dropped or modified by the policies.
Note that we can bypass this copy when processing does not depend on RPC argument values (e.g., rate limits).
During ACL policy enforcement, the RPC is dropped if the \texttt{key} argument is contained in a blocklist. Note that if an RPC is dropped, any further processing logic is never executed (including marshalling operations).

Finally, at the end of the processing logic, the transport adapter engine executes. \sys currently supports two types of transport: TCP and RDMA. For TCP, \sys uses the standard, kernel-provided scatter-gather (\texttt{iovec}) socket interface. For RDMA, \sys uses the scatter-gather \texttt{verb} interface, allowing the NIC to directly interact with buffers on the shared (or private) memory heaps containing the RPC metadata and arguments. 
For both TCP and RDMA, \sys provides disjoint memory blocks to the transport layer directly, eliminating excessive data movements.\footnote{For RDMA, if the number of disjoint memory blocks exceeds the limit of NIC’s capability to encapsulate all blocks in one RDMA work request, \sys coalesces the data into a memory block before transmission. This is because sending a single work request (even with a copy) is faster than sending multiple smaller work requests on our hardware.}

\parab{Control: Shared-memory queues.}
To facilitate efficient communication between an application and the \sysservice, we use shared memory control queues. \sys allocates two unidirectional queues for sending and receiving requests from an application to the \sysservice. The requests contain RPC descriptors, which reference arguments on the shared memory heap. The \sysservice always copies the RPC descriptors applications put in the sending queue to prevent TOCTOU attacks. \sys provides two options to poll the queues: 1) busy polling, and 2) \texttt{eventfd}-based adaptive polling. In busy polling, both the application-side \sys library and the \sysservice busy poll on their ends of the queues. In the \texttt{eventfd} approach, the \sys library and the \sysservice sends event notifications after enqueuing to an empty queue. After receiving a notification, the queue is drained (performing the necessary work) before subsequently waiting on future events.
The \texttt{eventfd} approach saves CPU cycles when queues are empty.
Other alternative solutions may involve dynamically scaling up (or down) the number of threads used to busy poll by the \sysservice; however, we chose the \texttt{eventfd} approach for its simplicity.
In our evaluation, we use busy polling for RDMA and eventfd-based adaptive polling for TCP.

\parab{Memory management.}
We provide a memory allocator in the \sys library for applications to directly allocate RPC data structures to be sent on a shared memory heap. The allocator invokes the \sysservice to allocate shared memory regions on behalf of the application (similar to how a standard heap manager calls \texttt{mmap} or \texttt{sbrk} to allocate memory from an OS kernel).
We need to use a specialized memory allocator for RPC messages (and their arguments), since RPCs are shared between three entities: the application, the \sysservice, and the NIC. A memory block is safe to be reclaimed only when it will no longer be accessed by any entity.

We adopt a notification-based mechanism for memory management. On the sender side, the outgoing messages are managed by the \sys library within the application. On the receiver side, the incoming messages are managed by the \sysservice. When the application no longer accesses a memory block occupied by outgoing messages, the memory block will not be reclaimed until the library receives a notification from \sysservice that the corresponding messages are already sent successfully through the NIC (similar to how zero-copy sockets work in Linux).
Incoming messages are put in buffers on a separate read-only shared heap. The receiving buffers can be reclaimed when the application finishes processing (e.g., when the RPC returns). To support reclamation of receive buffers, the \sys library notifies the \sysservice when specific messages are no longer in use by the application. Notifications for multiple RPC messages are batched to improve performance. If the receiver application code wishes to preserve or modify the incoming data, it must make an explicit copy.
Although this differs from traditional RPC semantics, in our implementation of Masstree and DeathStarBench we found no examples
where the extra copy was necessary.

\parab{Cross-\datapath policy engines.}
\sys supports engines that operate over multiple \datapaths, which may span multiple applications.
For instance, any global policy (e.g., QoS) will need to operate over all \datapaths (see \autoref{sec:advanced}).
For this type of engine, we instantiate replicas of the engine for each \datapath that it applies to.
Replicas can choose to either communicate through shared state, which requires managing contention across runtimes, or support runtime-local state that is contention-free.

\subsection{Live Upgrades}
\label{sec:upgrade}
Although our modular engine design for the \sysservice is similar to Snap~\cite{marty2019snap} and Click~\cite{kohler2000click}, we arrive at very different designs for upgrades. Click does not support live upgrades, while Snap executes the upgraded process to run alongside the old process. The old process serializes the engine states, transfers them to the new process, and the new process restarts them. This means that even changing a single line of code within a single Snap engine requires a complete restart for all Snap engines. This design philosophy is fundamentally not compatible with \sys, as we need to deal with new applications arriving with different RPC schemas, and thus our upgrades are more frequent. 
In addition, we want to avoid fate sharing for applications: changes to an application's \datapath should not impact the performance of other applications.
Ultimately, Snap is a network stack that does not contain application-specific code, where as \sys needs to be application-aware for marshalling RPCs.

We implement engines as plug-in modules that are dynamically loadable libraries.
We design a live upgrade method that supports \textit{upgrading, adding, or removing components of the \datapath without disrupting other \datapaths}.

\parab{Upgrading an engine.}
To upgrade one engine, \sys first detaches the engine from its runtime (preventing it from being scheduled).
Next, \sys destroys and deallocates the old engine, but maintains the old engine's state in memory; note that the engine is detached from its queues and not running at this time.
Afterwards, \sys loads the new engine and configures its send and receive queues. The new engine starts with the old engine's state. If there is a change in the data structures of the engine's state, the upgraded engine is responsible for transforming the state as necessary (which the engine developer must implement). 
Note that this also applies to any shared state for cross-\datapath engines.
The last step is for \sys to attach the new engine to the runtime.

\parab{Changing the \datapath.}
When an operator changes the \datapath to add or remove an engine, this process now involves the creation (or destruction) of queues and management of in-flight RPCs.
Changes that add an engine are straightforward, since it only involves detaching and reconfiguring the queues between engines.
Changes that remove an engine are more complex, as some in-flight RPCs may be maintained in internal buffers; for example, a rate limiter policy engine maintains an internal queue to ensure that the output queue meets a configured rate.
Engine developers are responsible for flushing such internal buffers to the output queues when the engines are removed.

\parab{Multi-host upgrades or \datapath changes.}
Some engine upgrades or \datapath changes that involve both the sender and the receiver hosts need to carefully manage in-flight RPCs across hosts. For example, if we want to upgrade how \sys uses RDMA, both the sender and the receiver have to be upgraded. In this scenario, the operator has to develop an upgrade plan that may involve upgrading an existing engine to some intermediate, backward-compatible engine implementation. The plan also needs to contain the upgrade sequence, e.g., upgrading the receiver side before the sender side.
Our evaluation demonstrates such a complex live upgrade, which optimizes the handling of many small RPC requests over RDMA (see \autoref{sec:upgrade-eval}).

\subsection{Security Considerations}
\label{sec:design-sec}
We envision two deployment models for \sys: (1) a cloud tenant uses \sys to manage its RPC workloads (similar to how sidecars are used today); (2) a cloud provider uses \sys to manage RPC workloads on behalf of tenants.
In both models, there are two different classes of principals: operators and applications.
Operators are responsible for configuring the hardware/virtual infrastructure,
deploying the mRPC service, and setting up policies that mRPC will enforce.
Applications run on an operator's infrastructure, interacting with the mRPC
service to invoke RPCs.
Applications trust operators, along with all privileged software (e.g., OS)
and hardware that the operators provide; both applications and operators
trust our mRPC service and protocol compiler.
In both deployment models, applications are not trusted and may be malicious (e.g., attempt to circumvent network policies).

In the first deployment model, \sysservice runs on top of a virtualized network that is dedicated to the tenant. Running arbitrary policy and observability code inside the \sysservice cannot attack other tenants' traffic since inter-tenant isolation is provided by the cloud provider.
In the second deployment model, our current prototype does not support running tenant-provided policy implementation inside \sysservice. How to safely integrate tenant-provided policy implementation and a cloud provider's own policy implementation is a future work.

From the application point of view, we want to ensure that \textbf{\sys provides equivalent security guarantees as compared to today's RPC library and sidecar approach}, which we discuss in terms of: 1) dynamic binding and 2) policy enforcement.
Our dynamic binding approach involves the generation, compilation, and
runtime loading of a shared library for (un)marshalling application RPCs.
Given that the compiled code is based on the application-provided RPC schema, this is a possible vector of attack.
The mRPC schema compiler is trusted with a minimal interface: other than providing the RPC schema, applications have no control on the process of how the marshalling code is generated.
We open source our implementation of the compiler so that it can be publicly reviewed.

As for all of our RPC processing logic, policies are enforced over RPCs by
operating over their representations in shared memory control queues and
data buffers.
With a naive shared memory implementation, this introduces a vector of
attack by exploiting a time-of-check to time-of-use (TOCTOU) attack; for
instance, the application could modify the RPC message after policy
enforcement but before the transport engine handles it.
In mRPC, we address this by copying data into an mRPC-private heap prior to
executing any policy that operates over the content of an RPC (as opposed to
metadata such as the length).
Similarly, received RPCs cannot be placed in shared memory until all
policies have been enforced, since otherwise applications could see received RPCs before policies have a chance to drop (or modify) them. 
Shared memory regions are maintained by the \sysservice on a per-application basis to provide isolation.

\section{Advanced Manageability Features}
\label{sec:advanced}

\sys's architecture creates an opportunity for advanced manageability features such as cross-application RPC scheduling. In this section, we present two such features that we developed on our policy engine framework to demonstrate the broader utility of our RPC-as-a-managed-service architecture.

\parab{Feature 1: Global RPC QoS.} 
\sys allows centralized RPC scheduling of cross-application workloads based on a global view of current outstanding RPCs. For example, \sys can enforce a policy that prioritizes RPCs with earliest deadlines~\cite{spuri1994efficient} across applications to support latency SLO or prioritizes latency-sensitive workloads~\cite{zhang2022justitia}.
One challenge here is that a naive implementation may attempt to apply the QoS policy for \datapaths spread over multiple runtimes (i.e., execution thread contexts).
This would require the (replicated) policy engines on each datapath to share the state on outstanding RPCs, and thus impose synchronization overheads.
Therefore, we adopt a similar strategy as used in the Linux kernel to apply the QoS policy on a per-runtime basis, which instead can use runtime-local storage without the need for synchronization.
In our implementation, we support a QoS strategy that prioritizes small RPCs based on a configurable threshold size.

\parab{Feature 2: Avoiding RDMA performance anomalies.}
It is well known that RDMA workloads may not fully utilize the capability of a specific RDMA NIC without fine-tuning, and that particular traffic patterns can even cause performance anomalies~\cite{kalia2016design, kong2022collie} (e.g., low RDMA throughput, pause frame storms). Previous work such as ScaleRPC~\cite{chen2019scalerpc} and Flock~\cite{monga2021flock} have proposed techniques to utilize the RNIC more efficiently. However, their approaches are library-based and only work for single applications; therefore, they do not handle scenarios in which the \emph{combination} of multiple application workloads causes poor RDMA performance. \sys's architecture enables us to have a global view of all RDMA requests and to avoid such performance anomalies.

We implement a global RDMA scheduler inside the RDMA transport engine, which translates RPC requests into RDMA messages and sends them to the RDMA NIC.
In our implementation, we focus on addressing the performance degradation from interspersed small and large scatter-gather elements (which may be across RPCs as well as applications).
We fuse such elements together with an explicit copy with an upper bound of 16\,KB for the size of the fused element.
\section{Implementation} \label{sec:impl}
\sys is implemented in 32K lines of Rust: 3K lines for the protocol compiler, 6K for the \sys control plane, 12K for engine implementations, and 11K for the \sys library. The \sys control plane is part of the \sysservice that loads/unloads engines. 

The \sys control plane is not live-upgradable. The \sys library is linked into applications and is thus also not live-upgradable. We do not envision the need to frequently upgrade these components because they only implement the high-level, stable APIs, such as shared memory queue communication and (un)loading engines. 

\newcommand{\apifunc}[1]{\texttt{#1}}

\begin{table}[t]
\small
\centering
\begin{tabular}{l}
\textbf{Operations} \\
\hline
\apifunc{doWork}(in:[Queue], out:[Queue]) \\
\hspace{2mm} \textit{Operate over one or more RPCs available on input queues.} \\
\hline
\apifunc{decompose}(out:[Queue]) $\rightarrow$ State \\
\hspace{2mm} \textit{Decompose the engine to its compositional states.} \\
\hspace{2mm} \textit{(Optionally output any buffered RPCs)}  \\
\hline
\apifunc{restore}(State) $\rightarrow$ Engine \\
\hspace{2mm} \textit{Restore the engine from the previously decomposed state.} \\
\end{tabular}
\vspace{-2mm}
\caption{\sys Engine Interface.}
\vspace{-5mm}
\label{tab:engine-api}
\end{table}

\parab{Engine interface.}
\autoref{tab:engine-api}~presents the essential API functions that all engines must implement.
Each engine represents some asynchronous computation that operates over input and output queues via \apifunc{doWork}, which is similar in nature to Rust's \texttt{Future}.
\sys uses a pool of runtime executors to drive the engines by calling \apifunc{doWork}, where each runtime executor corresponds to a kernel thread. We currently implement a simple scheduling strategy inspired by Snap~\cite{marty2019snap}: engines can be scheduled to a dedicated or shared runtime on start. In addition, runtimes with no active engines will be put to slept and release CPU cycles.
The engines also implement APIs to support live upgrading: \apifunc{decompose} and \apifunc{restore}. 
In \apifunc{decompose}, the engine implementation is responsible for destructing the engine and creating a representation of the final state of the engine in memory, returning a reference to \sys.
\sys invokes \apifunc{restore} on the upgraded instance of the engine, passing in a reference to the final state of the old engine.
The developer is responsible for handling backward compatibility across engine versions, similar to how application databases may be upgraded across changes to their schemas.

\parab{Transport engines.}
We abstract reliable network communication of messages into transport engines, which share similar design philosophy with Snap~\cite{marty2019snap} and TAS~\cite{kaufmann2019tas}. We currently implement two transport engines: RDMA and TCP. Our RDMA transport engine is implemented based on OFED libibverbs 5.4, while our TCP transport engine is built on Linux kernel's TCP socket.

\paragraph{\sys Library.}
Modern RPC libraries allow the user to specify the RPC data types and service interface through a language-independent schema file (e.g., \texttt{protobuf} for gRPC, \texttt{thrift} for Apache Thrift). \sys implements support for \texttt{protobuf} and adopts similar service definitions as gRPC, except for gRPC's streaming API.
\sys also integrates with Rust's async/await ecosystem for ease of asynchronous programming in application development.

To create an RPC service, the developer only needs to implement the functions declared in the RPC schema. The dependent RPC data types are automatically generated and linked with the application by the \sys schema compiler. The \sys library handles all the rest, including task dispatching, thread management, and error handling.
To allow applications to directly allocate data in shared memory without changing the programming abstraction, we implement a set of shared memory data structures that expose the same rich API as Rust's standard library. This is done by replacing the memory allocation of data structures such as \texttt{Vec} and \texttt{String} with the shared memory heap allocator. 
\section{Evaluation} \label{sec:eval}

We evaluate \sys using an on-premise testbed of servers with two 100 Gbps Mellanox Connect-X5 RoCE NICs and two Intel 10-core Xeon Gold 5215 CPUs (running at 2.5 GHz base frequency). The machines are connected via a 100 Gbps Mellanox SN2100 switch. Unless specified otherwise, we keep a single in-flight RPC to evaluate latency. To benchmark goodput and RPC rate, we let each client thread keep 128 concurrent RPCs on TCP and 32 concurrent RPCs on RDMA.

\subsection{Microbenchmarks} \label{sec:micro-bench}
We first evaluate \sys's performance through a set of microbenchmarks over two machines, one for the client and the other for the server. The RPC request has a byte-array argument, and the response is also a byte array. We adjust the RPC size by changing the array length. RPC responses are an 8-byte array filled with random bytes. We compare \sys with two state-of-the-art RPC implementations, eRPC and gRPC (v1.48.0). We deploy Envoy (v1.20) in HTTP mode to serve as a sidecar for gRPC. We use mRPC's TCP and RDMA backends to compare with gRPC and eRPC, respectively.
There is no existing sidecar that supports RDMA. To evaluate the performance of using a sidecar to control eRPC traffic, we implement a single-thread sidecar proxy using the eRPC interface. We keep applications running for 15 seconds to measure the result.

\parab{Small RPC latency.}
We evaluate \sys's latency by issuing 64-byte RPC requests over a single connection. \autoref{tab:micro-latency} shows the latency for small RPC requests. Note that since the marshalling of small messages is fast on modern CPUs, the result in the table remains stable even when the message size scales up to 1\,KB. We use \texttt{netperf} and \texttt{ib\_read\_lat} to measure raw round-trip latency.

\sys achieves median latency of 32.8\,\us for TCP and 7.6\,\us for RDMA. Relative to netperf (TCP) or a raw RDMA read, \sys adds 11.8 or 5.1\,\us to the round-trip latency. 
This is the cost of the \sys abstraction on top of the raw transport interface (e.g., socket, verbs).

We also evaluate latency in the presence of sidecar proxies. 
The sidecars do not enforce any policies, so we are only measuring the base overhead. Our results show that adding sidecars substantially increases the RPC latency. On gRPC, adding Envoy sidecars more than triples the median latency. The result is similar with eRPC. On mRPC, having a NullPolicy engine (which simply forwards RPCs) in the \sysservice has almost no effect on latency, increasing the median latency only by 300\,ns.

Comparing the full solution (mRPC with policy versus gRPC/eRPC with proxy), mRPC speeds up the median latency by 6.1$\times$ (i.e., 33.4\,\us against 203.4\,\us) and the 99th percentile tail latency by 5.8$\times$. On RDMA, \sys speeds up eRPC by 1.3$\times$ and 1.4$\times$ in terms of median and tail latency (respectively). This is because the communication between the eRPC app and its proxy goes through the NIC, which triples the cost in the end-host driver (including the PCIe latency). In contrast, \sys's architecture shortcuts this step with shared memory.

In addition, to separate the performance gain from system implementation difference, we evaluate the latency of mRPC with full gRPC-style marshalling (protobuf encoding and HTTP/2 framing) in the presence of NullPolicy engines as an ablation study. Under this setting, compared with gRPC + Envoy, \sys speeds up the latency by 4.1$\times$ in terms of both median and tail latency.
We also observe that the mRPC framework does not introduce significant overhead. Even with the cost of protobuf and HTTP/2 encoding, mRPC still achieves slightly lower latency compared with standalone gRPC. In \sys, we can choose a customized marshalling format, because we know the other side is also an \sysservice. In other cases, e.g., when interfacing with external traffic or dealing with endianness differences, we can still apply full-gRPC style marshalling. 
When mRPC is configured to use full-gRPC style marshalling, we only need to pay (un)marshalling costs between mRPC services. For gRPC + Envoy, in addition to the (un)marshalling costs between Envoy proxies, the communication between applications and Envoy proxies also needs to pay this (un)marshalling cost. 
In the remaining evaluations, we will use \sys's customized marshalling protocol. More results using gRPC-style marshalling are shown in \autoref{sec:eval-mrpc-proto}.

\begin{table}[t]
\centering
\resizebox{0.48\textwidth}{!}{%
\begin{tabular}{cccc}
\hline
\multicolumn{1}{l}{\textbf{Transport}} &
  \multicolumn{1}{l}{\textbf{\space \space \space \space \space \space \space \space \space \space Solution}} &
  \multicolumn{1}{l}{\textbf{Median Latency (\us)}} &
  \multicolumn{1}{l}{\textbf{P99 Latency (\us)}} \\ \hline
\multicolumn{1}{c|}{\multirow{5}{*}{TCP}}  & Netperf   & 21.0 & 32.0 \\ \cline{2-4}
\multicolumn{1}{c|}{}                      & gRPC      & 63.0 & 90.3 \\
\multicolumn{1}{c|}{}                      & mRPC      &  32.8 & 38.7 \\
\multicolumn{1}{c|}{}                      & gRPC+Envoy & 203.4 & 251.1 \\ 
\multicolumn{1}{c|}{}                      & mRPC+NullPolicy & \textbf{33.4} & \textbf{43.3} \\ 
\multicolumn{1}{c|}{}                      & mRPC+NullPolicy+HTTP+PB & 49.8 & 61.9 \\ \hline
\multicolumn{1}{c|}{\multirow{5}{*}{RDMA}} & RDMA read & 2.5  & 2.8  \\ \cline{2-4}
\multicolumn{1}{c|}{}                      & eRPC      & 3.6  & 4.1  \\
\multicolumn{1}{c|}{}                      & mRPC      & 7.6  & 8.7  \\
\multicolumn{1}{c|}{}                      & eRPC+Proxy & 11.3  & 15.6  \\
\multicolumn{1}{c|}{}                      & mRPC+NullPolicy & \textbf{7.9}  & \textbf{9.1}  \\ \hline
\end{tabular}%
}
\vspace{-2mm}
\caption{\textbf{Microbenchmark [Small RPC latency]:} Round-trip RPC latencies for 64-byte requests and 8-byte responses.}
\vspace{-4mm}
\label{tab:micro-latency}
\end{table}

\begin{figure}[t]
\centering
\vspace{-0mm}
\subfloat[][TCP-based transport]{%
    \label{fig:bench_bw_tcp}
    \includegraphics[scale=0.28]{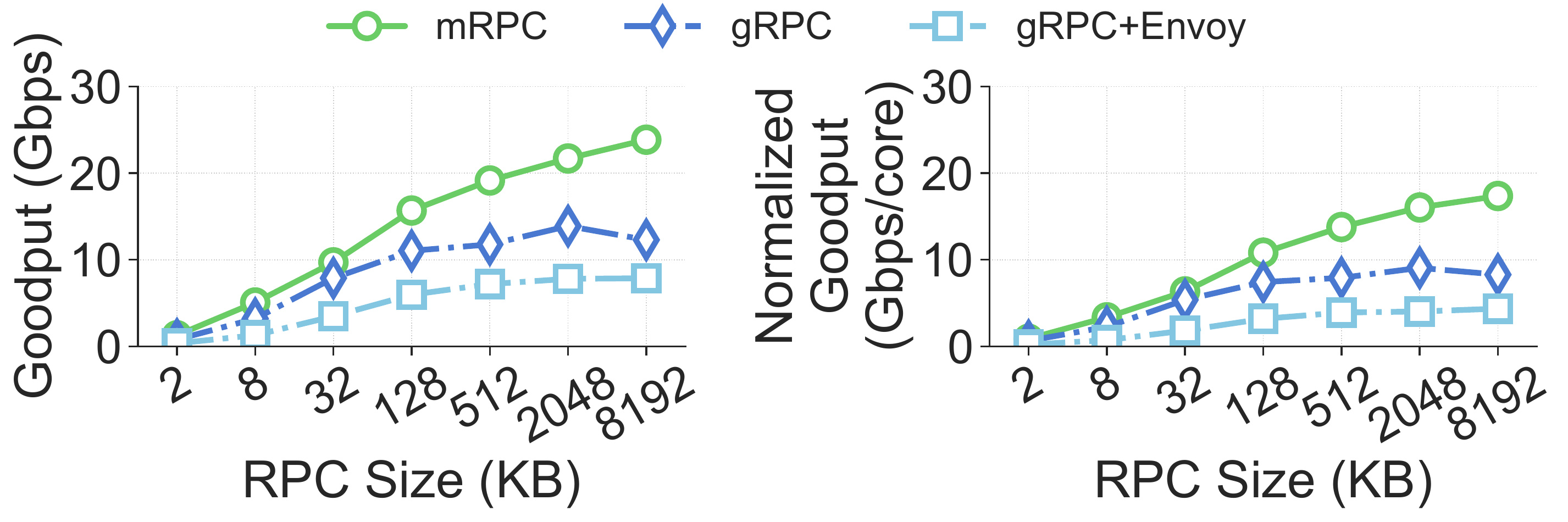}
}
\hfill
\subfloat[][RDMA-based transport]{%
    \label{fig:bench_bw_rdma}
    \includegraphics[scale=0.28]{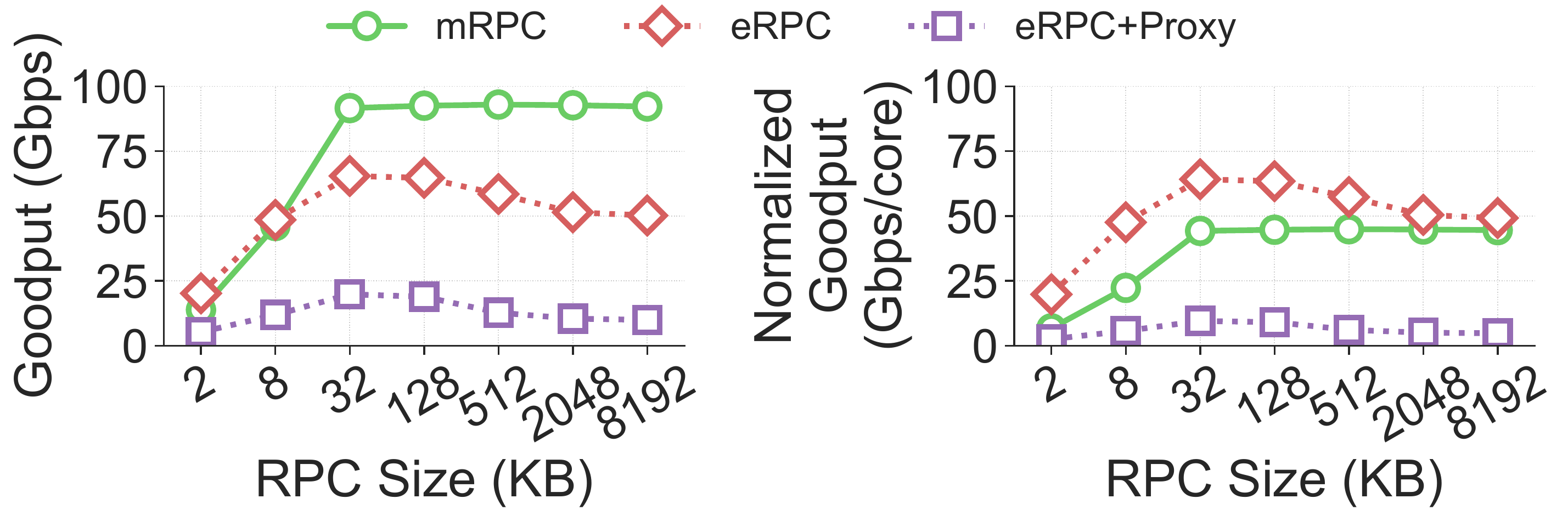}
}
\vspace{-0mm}
\caption[]{\textbf{Microbenchmark [Large RPC goodput]:} Comparison of goodput for large RPCs. Note that different solutions demand different amounts of CPU cores, so we also normalized the goodput to their CPU utilization, as shown in the right figures. The error bars show the 95\% confidence interval, but they are too small to be visible.}
\vspace{0mm}
\label{fig:micro-bandwidth}
\end{figure}

\parab{Large RPC goodput.}
The client and server in our goodput test use a single application thread.
The left side of \autoref{fig:micro-bandwidth} shows the result. From this point on, when we discuss \sys's performance, we focus on the performance of \sys that has at least a NullPolicy engine in place to fairly compare with sidecar-based approaches.

\sys speeds up gRPC + Envoy and eRPC + Proxy, by 3.1$\times$ and 9.3$\times$, respectively, for 8KB RPC requests. \sys is especially efficient for large RPCs\footnotemark, for which (un)marshalling takes a higher fraction of CPU cycles in the end-to-end RPC datapath. Having a sidecar substantially hurts RPC goodput both for TCP and RDMA. In particular, for RDMA, intra-host roundtrip traffic through the RNIC might contend with inter-host traffic in the RNIC/PCIe bus, halving the available bandwidth for inter-host traffic. \sys even outperforms gRPC (without Envoy). \sys is fundamentally more efficient in terms of marshalling format: \sys uses \texttt{iovec} and incurs no data movement. \autoref{sec:eval-mrpc-proto} shows an ablation study that demonstrates that even if \sys uses a full gRPC-style marshalling engine, \sys outperforms gRPC + Envoy due to a reduction in the number of (un)marshalling steps.

\parab{CPU overheads.}
To understand the \sys CPU overheads, we measure the per-core goodput. The results are shown on the right side of \autoref{fig:micro-bandwidth}. \sys speeds up gRPC + Envoy and eRPC + Proxy, by 3.8$\times$ and 9.3$\times$, respectively. This means \sys is much more CPU-efficient than gRPC + Envoy and eRPC + Proxy. eRPC (without a proxy) is quite efficient, but converges to mRPC's efficiency as RPC size increases. \footnotetext{Standalone eRPC exhibits relatively lower goodput on RoCE than on Infiniband. According to the eRPC paper~\cite{kalia2019erpc}, eRPC should achieve 75\,Gbps on Infiniband for 8MB RPCs.}

\begin{figure}[t]
\centering
\vspace{0mm}
\subfloat[][TCP-based transport]{%
    \label{fig:bench_rate_tcp}
    \includegraphics[scale=0.28]{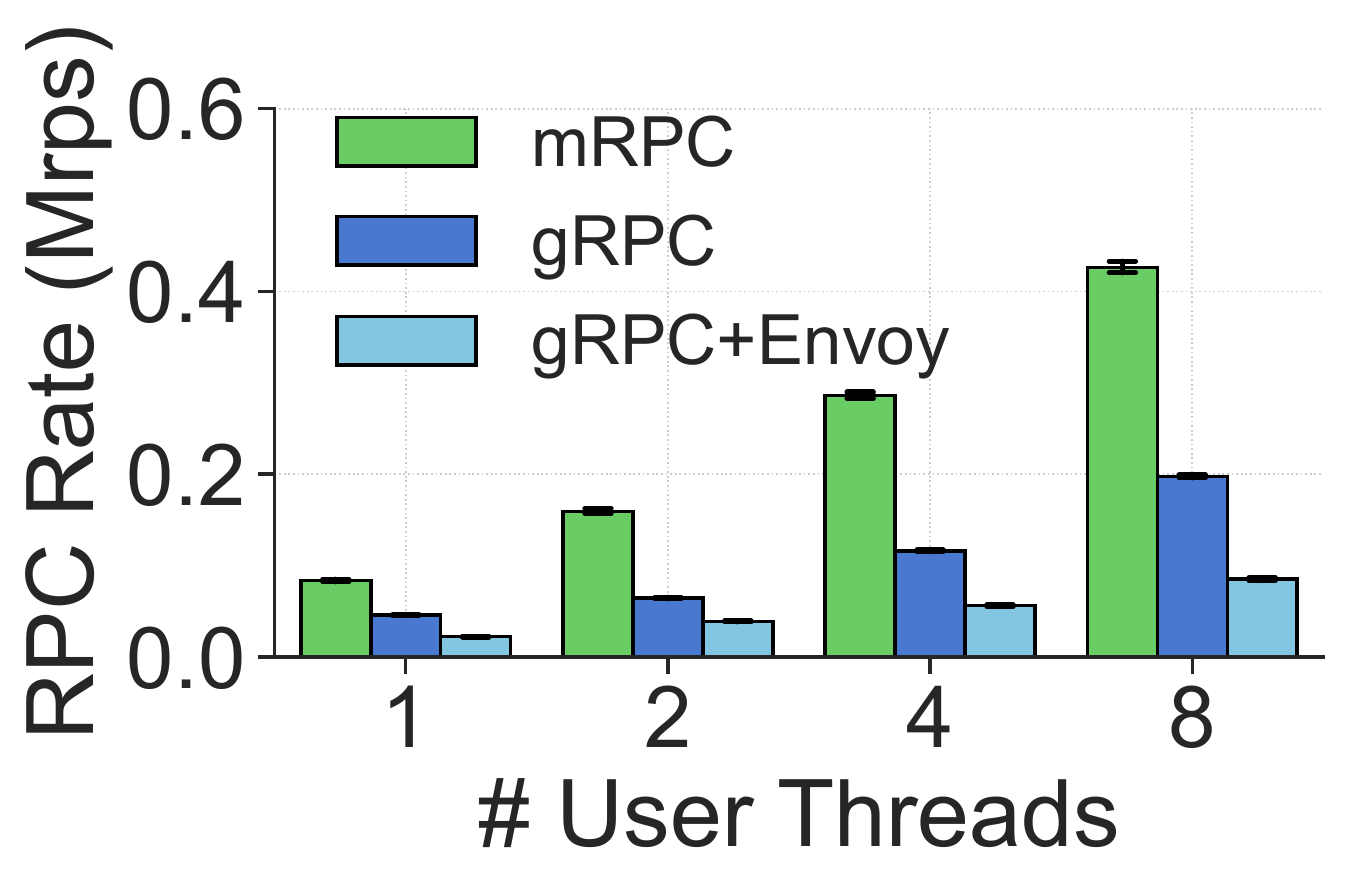}
}
\hfill
\subfloat[][RDMA-based transport]{%
    \label{fig:bench_rate_rdma}
    \includegraphics[scale=0.28]{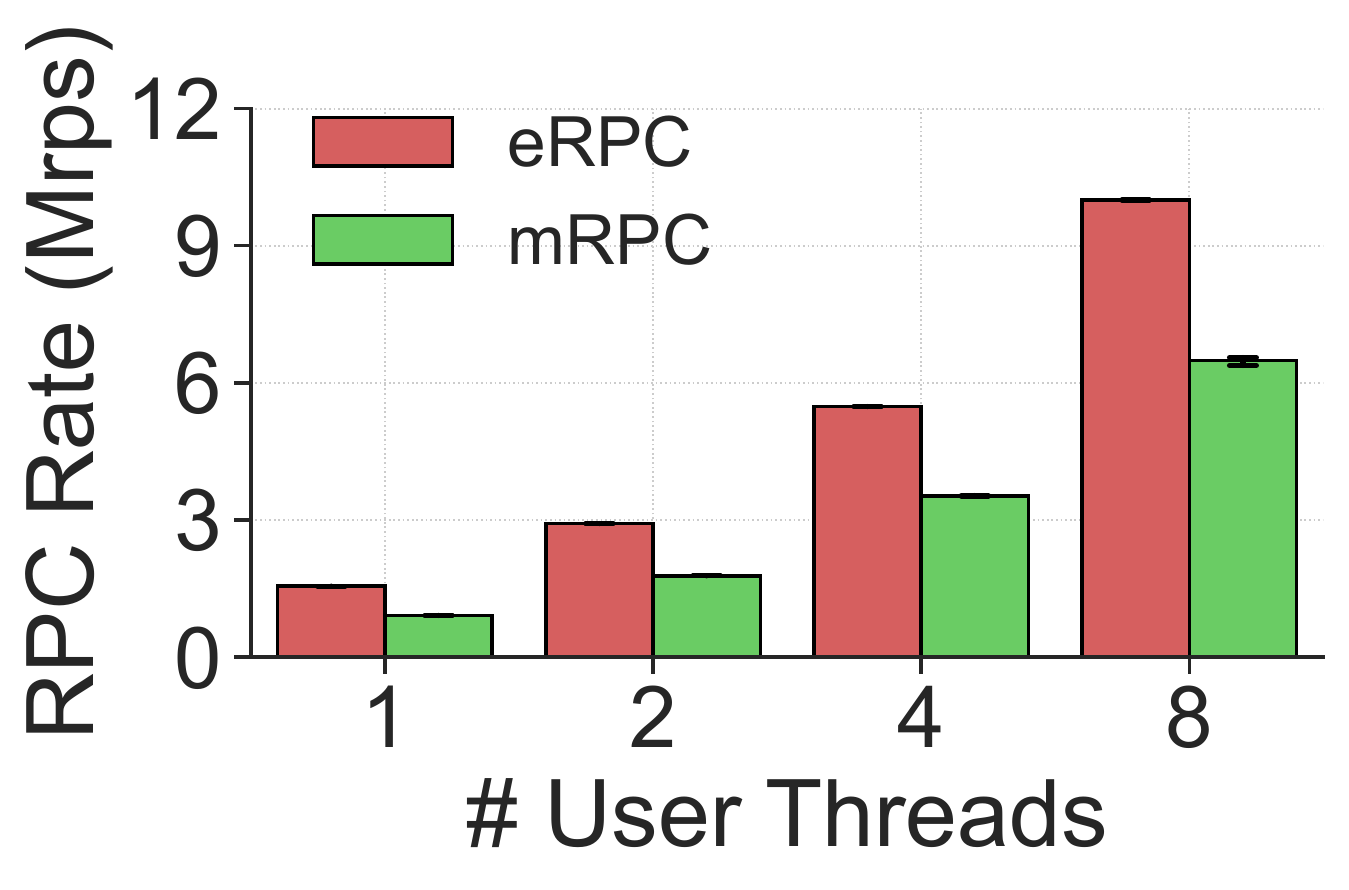}
}
\vspace{0mm}
\caption{\textbf{Microbenchmark [RPC rate and scalability]:} Comparison of small RPC rate and CPU scalability. The bars show the RPC rate. The error bars show the 95\% confidence interval.}
\vspace{-2mm}
\label{fig:micro-rate}
\end{figure}

\parab{RPC rate and scalability.}
We evaluate \sys's small RPC rate and its multicore scalability. We fix the RPC request size to 32 bytes and scale the number of client threads.
We use the same number of threads for the server as the client, and each client connects to one server thread. \autoref{fig:micro-rate} shows the RPC rates when scaling from 1 to 8 user threads. All the tested solutions scale well. \sys's RPC rates scale by 5.1$\times$ and 7.2$\times$, on TCP and RDMA, from a single thread to 8 threads. As a reference, gRPC scales by 4.3$\times$, gRPC + Envoy scales by 3.9$\times$, and eRPC scales by 6.5$\times$. \sys achieves 0.43\,Mrps on TCP and 6.5\,Mrps on RDMA with 8 threads. gRPC + Envoy only has 0.09\,Mrps, so \sys outperforms it by 5$\times$. We do not evaluate eRPC + proxy, because our eRPC proxy is only single-threaded. When we run eRPC + proxy with a single thread, it achieves 0.51\,Mrps. So even if eRPC + proxy scales linearly to 8 threads, \sys still outperforms it.

\subsection{Efficient Policy Enforcement}
\label{sec:policy}
We use two network policies as examples to demonstrate \sys's efficient support for RPC policies: (1) RPC rate limiting and (2) access control based on RPC arguments. RPC rate limiting allows an operator to specify how many RPCs a client can send per second. We implement rate limiting as an engine using the token bucket algorithm~\cite{tokenbucket}. Our access control policy inspects RPC arguments and drops RPCs based on a set of rules specified by network operators. These two network policies differ greatly from traditional rate limiting and access control, which only limit network bandwidth and can only operate on packet headers.

\begin{figure}
	\begin{subfigure}{0.45\linewidth}
      \includegraphics[width=\linewidth]{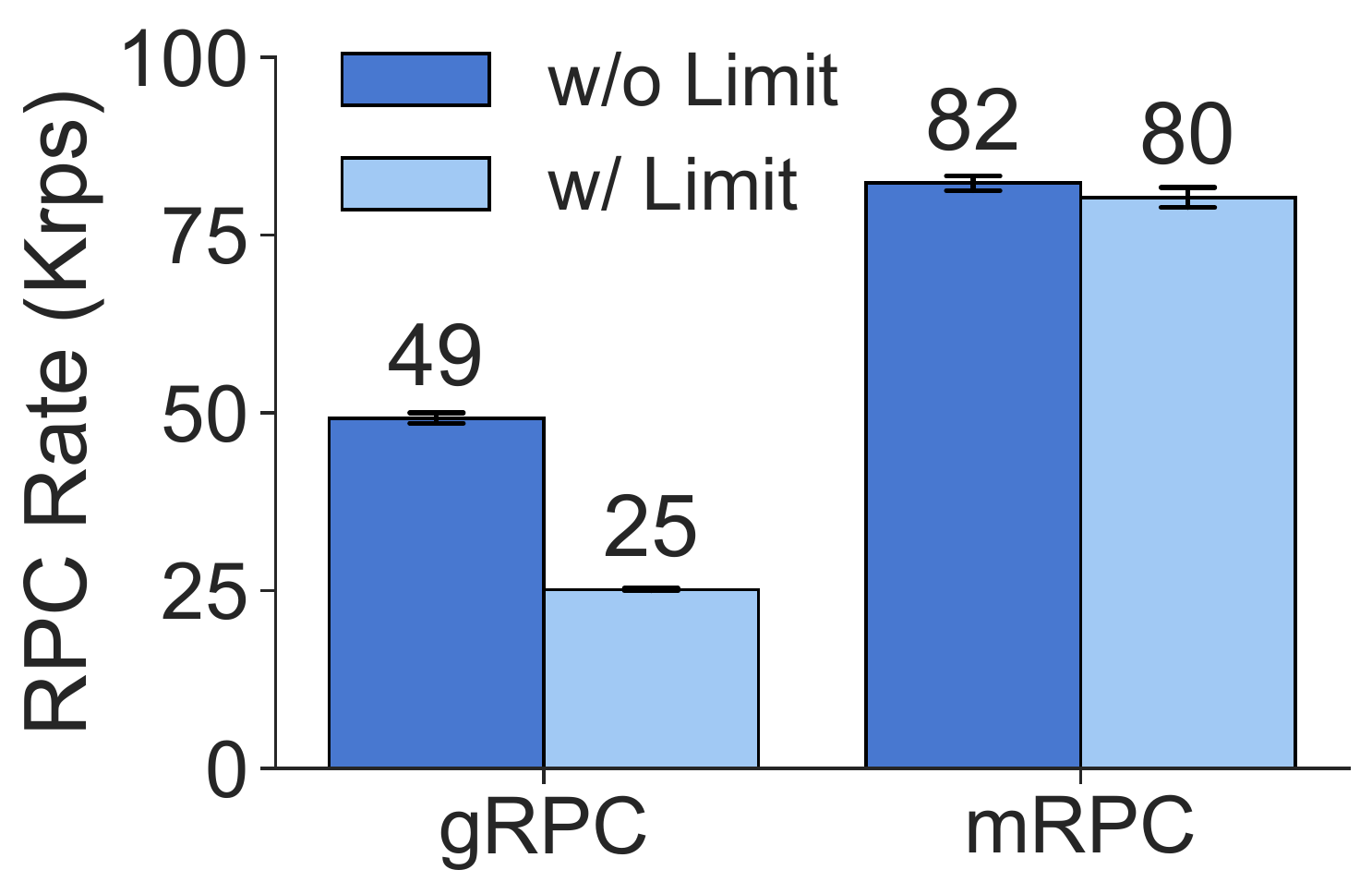}
       \caption{Rate Limiting}
       \label{fig:ratelimit_bar}
    \end{subfigure} \hfil
	\begin{subfigure}{0.45\linewidth}
      \includegraphics[width=\linewidth]{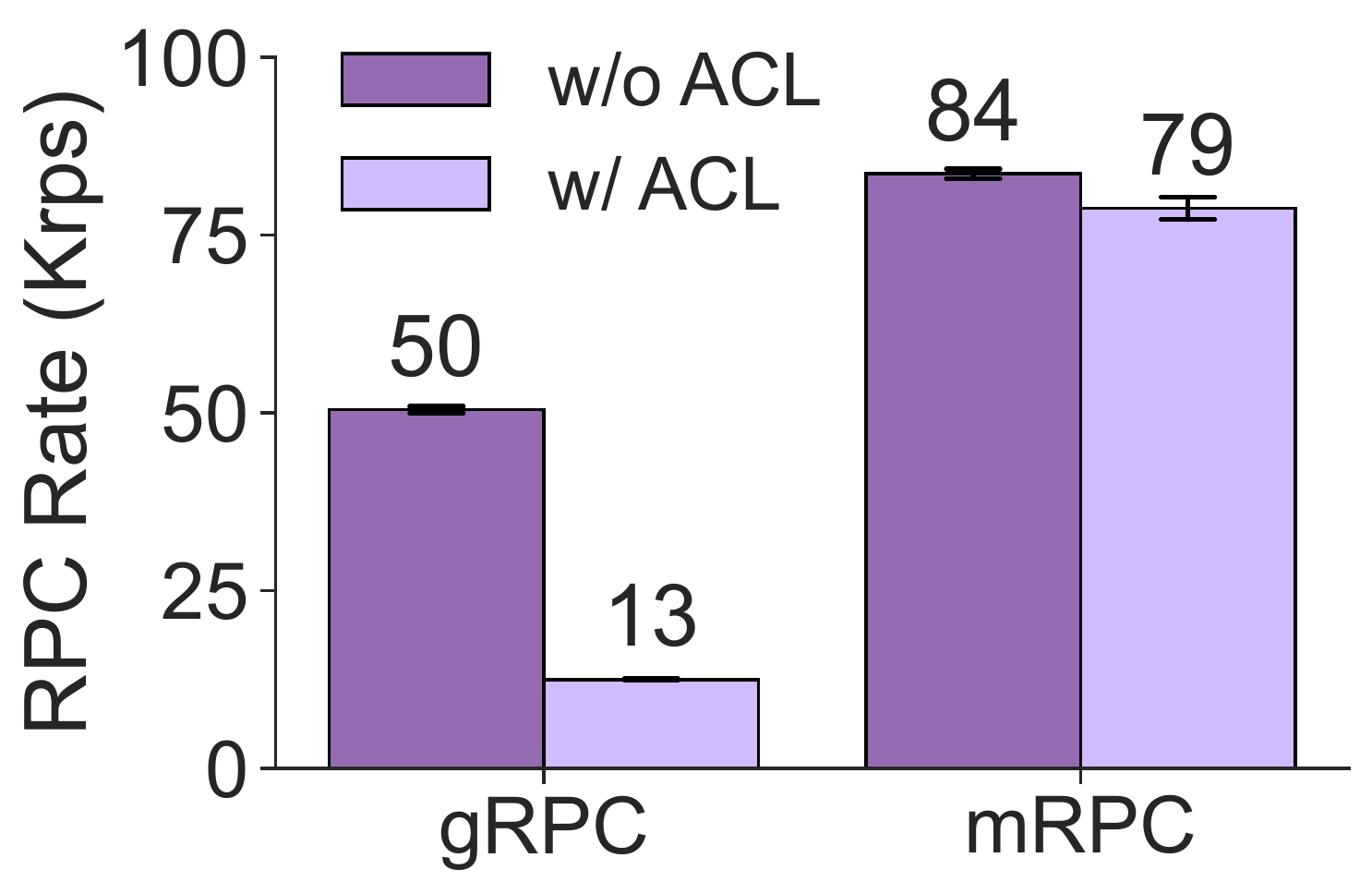}
        \caption{Access Control}
        \label{fig:acl_bar}
    \end{subfigure}
\vspace{-2mm}
\caption{\textbf{Efficient Support for Network Policies}. The RPC rates with and without policy are compared. The bars of w/o Limit and w/o ACL for gRPC show its throughput when the sidecar is bypassed. The error bars show the 95\% confidence interface.} 
\vspace{-2mm}
\label{fig:policy}
\end{figure}

We compare rate limit enforcement using an mRPC policy versus using Envoy's rate limiter on gRPC workloads. To evaluate the performance overheads, we set the limit to infinity so that the actual RPC rate is never above the limit (allowing us to observe the overheads).
\autoref{fig:ratelimit_bar} shows the RPC rate with and without the rate limits. gRPC's RPC rate drops immediately from 49K to 25K. This is because having a sidecar proxy (Envoy) introduces substantial performance overheads. For mRPC, the RPC rate stays the same at 82K. This is because having a policy introduces minimal overheads. The extra policy only adds tens to hundreds of extra CPU instructions on the RPC datapath.

We evaluate access control on a hotel reservation application in DeathStarBench~\cite{gan2019deathstarbench}. The service handles hotel reservation RPC requests, which include the customer's name, the check-in date, and other arguments. The service then returns a list of recommended hotel names. We set the access control policy to filter RPCs based on the \texttt{customerName} argument in the request.
We use a synthetic workload containing 99\% valid and 1\% invalid requests. We again compare our \sys policy against using Envoy to filter gRPC requests.
We implement the Envoy policy using WebAssembly. gRPC's rate drops from 50K to 13K. This is because of the same sidecar overheads and now Envoy has to further parse the packets to fetch the RPC arguments. On mRPC, the performance drop is much smaller, from 84K to 79K. Note that, on mRPC, the performance overhead of introducing access control is larger than rate limiting. For access control, the \sysservice has to copy the relevant field (i.e., \texttt{customerName}) to the private heap to prevent TOCTOU attacks on the sender side and has to copy the RPC from a private heap to the shared heap on the receiver side.

\subsection{Live Upgrade}
\label{sec:upgrade-eval}

We demonstrate mRPC's ability to live upgrade using two scenarios.

\parab{Scenario 1.}
During our development of \sys, we realized that using the RDMA NIC's scatter-gather list to send multiple arguments in a single RPC can significantly boost \sys's performance. In this approach, even when an RPC contains arguments that are scattered in virtual memory, we can send the RPC using a single RDMA operation (\texttt{ibv\_post\_send}).
We use these two versions of our RDMA transport engine to demonstrate that \sys enables such an upgrade without affecting running applications.
Note that all other evaluations already include this RDMA feature.
This upgrade involves both the client side's \sysservice and the server side's \sysservice, because it involves how RDMA is used between machines (i.e., transport adapter engine). gRPC and eRPC cannot support this type of live upgrade.

\begin{figure}[t]
	\begin{subfigure}{0.45\linewidth}
      \includegraphics[width=\linewidth]{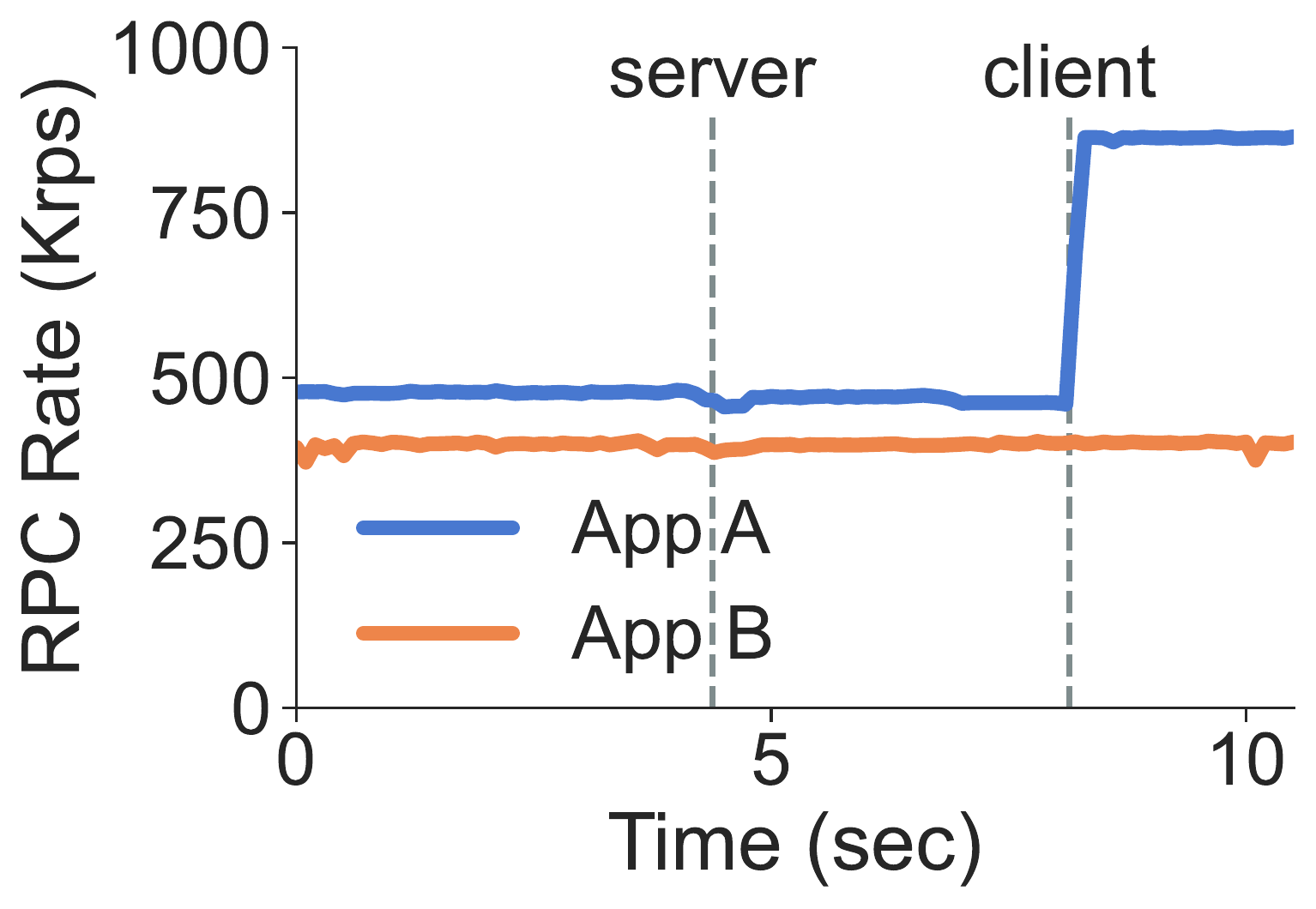}
       \caption{Transport adapter}
       \label{fig:live_upgrade_transport}
    \end{subfigure} \hfil
	\begin{subfigure}{0.45\linewidth}
      \includegraphics[width=\linewidth]{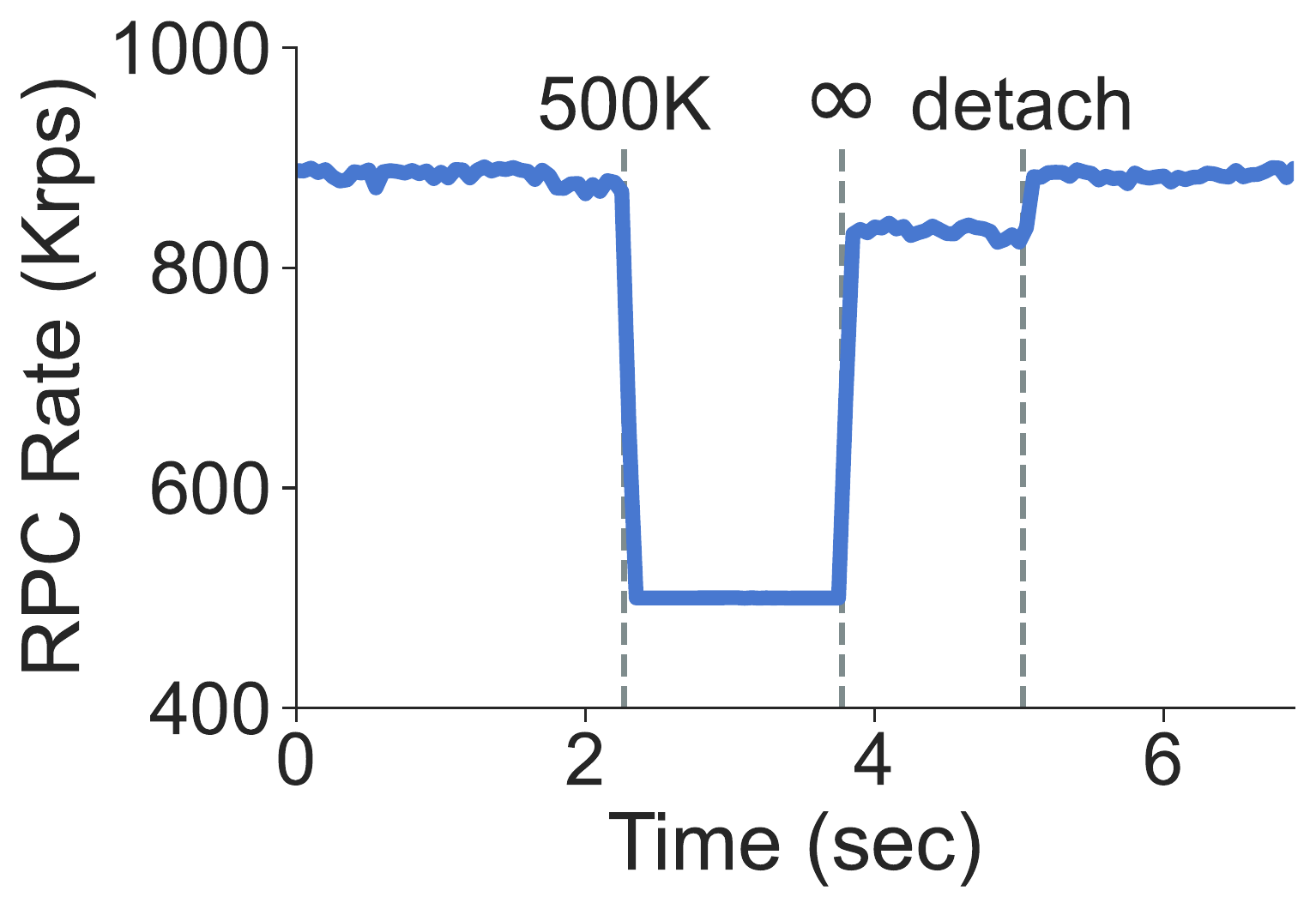}
        \caption{Rate limit policy}
        \label{fig:live_upgrade_policy}
    \end{subfigure}\vspace{-0.1in}
\caption{\textbf{Live upgrade}. In (a), the annotations indicate when the client 
 of App A and server of A and B are upgraded. In (b), the annotations denote the specified rate and when the policy is removed.}
\vspace{-0mm}
\label{fig:live_upgrade}
\end{figure}

We run two applications (App A and App B). Both applications are sending 32-byte RPCs, and the responses are 8 bytes. A and B share the \sysservice on the server side. A's and B's RPC clients are on different machines.
We keep 8 concurrent RPCs for B, forcing it to send at a slower rate, while using 32 for A.  We first upgrade the server side to accept arguments as a scatter-gather list, and we then upgrade the client side of A. \autoref{fig:live_upgrade_transport} shows the RPC rate of A and B.
When the server side upgrades, we observe a negligible effect on A's and B's rate. Neither A nor B needs recompilation or rebooting. When A's client side's \sysservice is upgraded, A's performance increases from 480K to 860K. B's performance is not affected at all because B's client side's \sysservice is not upgraded. 

\parab{Scenario 2.}
Enforcing network policies has performance overheads, even when they do not have any effect. For example, enforcing a rate limit of an extremely large throttle rate still introduces performance overheads just for tracking the current rate using token buckets. \sys allows policies to be removed at runtime, without disrupting running applications.

We use the same rate limiting setup from \autoref{sec:policy} but on top of RDMA transport. \autoref{fig:live_upgrade_policy} shows the RPC rate. We start from not having the rate limit engine. We then load the rate limit engine and set the throttled rate to 500K. The RPC rate immediately becomes 500K. We then set the throttled rate to be infinite, and the rate becomes 840K. After we detach the rate limit engine, the rate becomes 890K.

\parab{Takeaways.}
There are two overall takeaways from these experiments. First, \sys allows upgrades to the \sysservice without disrupting running applications. Second, live upgrades allow for more flexible management of RPC services, which can be used to enable immediate performance improvements (without redeploying applications) or dynamic configuration of policies.

\subsection{Real Applications}
\label{sec:real-apps}

We evaluate how the performance benefits of \sys transform into end-to-end application-level performance metrics.

\begin{figure}
\centering
\vspace{-0mm}
\includegraphics[width=0.8\columnwidth]{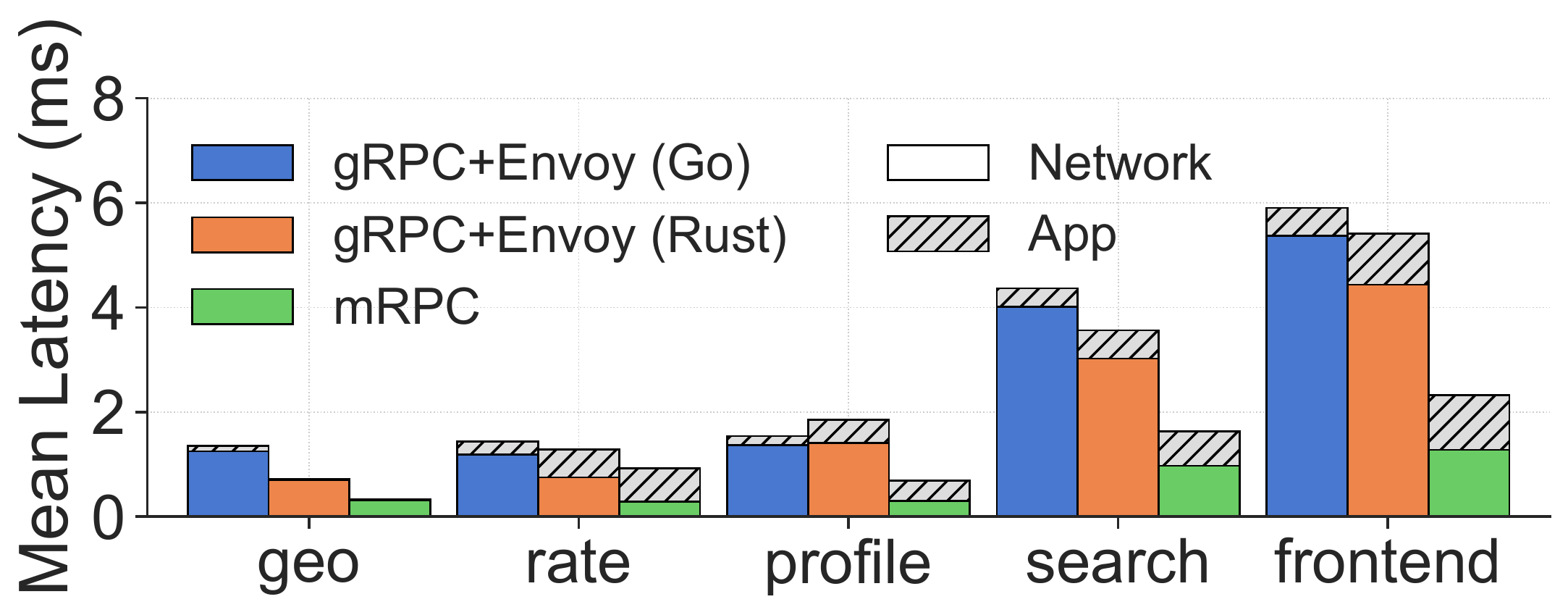}
\vspace{-0mm}
\caption{\textbf{DeathStarBench}: Mean latency of in-application processing and network processing of microservices. The latency of a microservice includes RPC calls to other microservices. The frontend latency represents complete end-to-end latency.}
\vspace{-0mm}
\label{fig:deathstarbench}
\end{figure}

\parab{DeathStarBench.}
We use the hotel reservation service from the DeathStarBench~\cite{gan2019deathstarbench} microservice benchmark suite.
The reference benchmark is implemented in Go with gRPC and Consul~\cite{consul} (for service discovery). Our \sys prototype currently only supports Rust applications, and we thus port the application code to Rust for comparison. We use the same open-source services such as memcached~\cite{memcached} and MongoDB~\cite{mongodb}.

We distribute the HTTP frontend and the microservices on four servers in our testbed. The monolithic services (memcached, MongoDB) are co-located with the microservices that depend on them. We use a single thread for each of the microservices and the frontend. Further, we deploy an Envoy proxy as a sidecar on each of the servers (with no active policy). The provided workload generator~\cite{gan2019deathstarbench} is used to submit HTTP requests to the frontend. For a fair comparison, we also implemented a Rust version of the benchmark with Tonic~\cite{tonic}, which is the de facto implementation of gRPC in Rust. 
We deploy the mRPC and Tonic implementations on bare metal, while the reference Go suite runs in Docker containers with a host network (which introduces negligible performance overheads compared to using bare metal~\cite{zhuo2019slim}).
All three solutions are based on TCP.
We issue 20 requests per second for 250 seconds and record the latency of each request, breaking it down into the in-application processing time and network processing time for each microservice involved. In our evaluation, the dynamic bindings of the user applications are already cached in \sysservice, so the time to generate the bindings is not included in the result.

\autoref{fig:deathstarbench} shows the latency breakdown. First, we validate that our own implementation of DeathStarBench on Rust is a faithful re-implementation. We can see that the original Go implementation and our Rust implementation have similar latency. Moreover, the amount of latency spent in gRPC is similar. Second, \sys with a null policy outperforms by 2.5$\times$ gRPC with a sidecar proxy in average end-to-end latency. \autoref{sec:real-app-appendix} contains more details about the tail latency and the scenario without a sidecar.

\begin{table}
    \centering
    \small
    \begin{tabular}{lccc}
        \toprule
        & Median Latency & P99 Latency & Throughput \\
        \midrule
        eRPC & 16.8 \us & 21.7 \us & 8.7 MOPS \\
        mRPC & 22.5 \us & 33.1 \us & 7.0 MOPS \\
        \bottomrule
    \end{tabular}
    \vspace{-0mm}
    \caption{\textbf{Masstree analytics}: Latency and the achieved throughput for GET operations. MOPS is Million Operations Per Second.}
    \vspace{-0mm}
    \label{tab:masstree}
\end{table}

\parab{Masstree analytics.}
We also evaluate the performance of Masstree~\cite{mao2012cache}, an in-memory key-value store, over both \sys and eRPC~\cite{kalia2019erpc} using RDMA. We follow the exact same workload setup used in eRPC, which contains 99\% I/O-bounded point GET request and 1\% CPU-bounded range SCAN request. We run the Masstree server on one machine and run the client on another machine. Both the server and the client use 10 threads, with each client thread using 16 concurrent requests. The test runs for 60 seconds.
The result in~\autoref{tab:masstree} shows that eRPC outperforms \sys, which makes sense since eRPC is a well-designed library implementation that is focused on high performance.
\sys enables many other manageability features in exchange for a slight reduction in performance. In this case, using \sys instead of eRPC means that median latency increases by 34\% and throughput reduces by 20\%.

\subsection{Benefits of Advanced Manageability Features}

\begin{table}
    \centering
    \small
    \begin{tabular}{lccc}
        \toprule
        & \multicolumn{2}{c}{Latency App} & B/W App \\
        \cmidrule(lr){2-4}
        & P95 Latency & P99 Latency & Bandwidth \\
        \midrule
        w/o QoS & 45.1 \us & 54.6 \us & 22.2 Gbps \\
        w/ QoS & 19.5  \us & 21.8 \us & 22.0 Gbps \\ 
        \bottomrule
    \end{tabular}
    \vspace{-0mm}
    \caption{\textbf{Global QoS}: Performance of latency- and bandwidth-sensitive applications with and without a global QoS policy.}
    \vspace{0mm}
    \label{tab:global_qos}
\end{table}

Next, we demonstrate the performance benefits of having centralized RPC management, through two advanced manageability features that we developed (see~\autoref{sec:advanced}). We use synthetic workloads to test the advanced manageability features.

\parab{Global RPC QoS.}
We enable our cross-application QoS policy that reorders requests from multiple applications and prioritizes small RPC quests.
We set up two applications and pin them to the same \sys runtime. One application is latency-sensitive,  sending 32-byte RPC requests with a single RPC in-flight; the other is bandwidth-sensitive, sending 32\,KB requests with 64 concurrent RPCs.
We measure the tail latency for the latency-sensitive application and the utilized bandwidth of the bandwidth-sensitive one.

\autoref{tab:global_qos} shows the result. Without the QoS policy, the bandwidth-sensitive application has a high bandwidth utilization; however, the latency-sensitive application suffers from a high tail latency. With the QoS policy in place, the small requests from the latency-sensitive application get higher priority and are sent first, improving P99 latency from 54.6 \us to 21.8 \us. Since small RPC requests consume negligible bandwidth, it barely affects the bandwidth-sensitive application (less than a 1\% bandwidth drop).

\begin{figure}
\centering
\includegraphics[width=0.8\columnwidth]{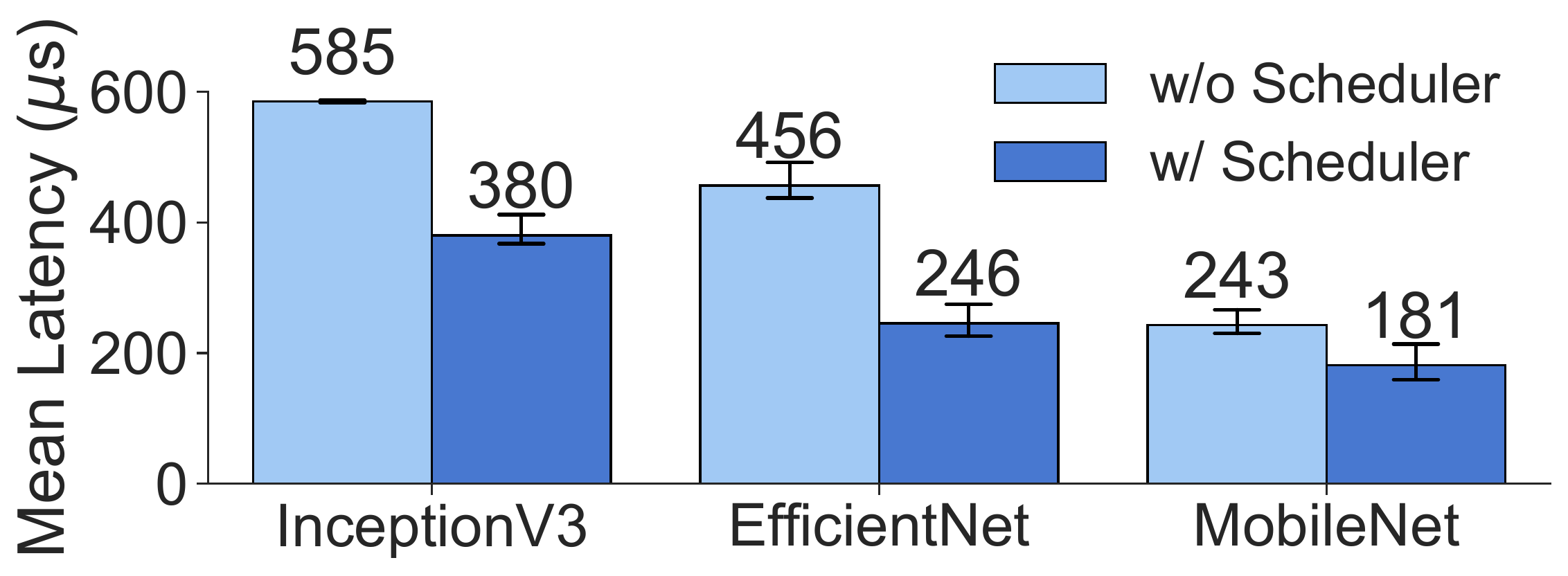}
\vspace{-0mm}
\caption{\textbf{RDMA Scheduler}: Mean RPC latency with or without RDMA scheduler. The error bars show the 95\% confidence interval.} 
\label{fig:scheduler-real-workload}
\vspace{-0mm}
\end{figure}

\parab{RDMA Scheduler.}
Our RDMA scheduler batches small RPC requests into (at most) 16KB messages and sends requests using a single RMDA operation to reduce the load on the RDMA NIC.
Our synthetic workload is based on BytePS~\cite{jiang2022byteps}, which uses RDMA for distributed deep learning.
To synchronize a tensor to/from a server, BytePS prepends an 8-byte key and appends a 4-byte length to describe the tensor. The three disjoint memory blocks are placed in a scatter-gather list and submitted to the NIC, resulting in a small-large-small message pattern that triggers a performance anomaly~\cite{kong2022collie}. This message pattern is quite common in real applications, as programs often need to describe a large payload with a small piece of metadata.
We emulate BytePS's RPC request pattern and generate RPCs from three widely-used models: MobileNet, EfficientNetB0, and InceptionV3~\cite{Howard2017MobileNet, tan2019efficientnet, szegedy2016rethinking}. Each RPC call consists of an 8-byte key, a payload of tensor, and a 4-byte length. We use a single thread to make RPCs. \autoref{fig:scheduler-real-workload} shows the average RPC latency. The RDMA scheduler provides 30-90\% latency improvement. This improvement differs for different neural networks, because of different RDMA message patterns. 

\section{Related Work} \label{sec:related}
\paragraph{Fast RPC implementations.}
Optimizing RPC has a long history. Birrell and Nelson's early RPC design~\cite{birrell1984rpc} includes generating bindings via a compiler, interfacing with transport protocols, and various optimizations (e.g., implicit ACK). Bershad et al. showed how to
use shared-memory queues to efficiently pass RPC messages between processes on the same machine~\cite{bershad1991urpc}.
\sys's shared-memory region leverages this idea but extends it to allow for marshalling code to be applied after
policy enforcement. A similar use of shared-memory queues can be found with recent Linux support for asynchronous system calls~\cite{iouring} combined with scatter-gather I/O~\cite{linuxlove}; unlike traditional system calls, however, \sys protocol 
descriptions can be defined at runtime. 

Another line of work uses RDMA to speed network RPCs~\cite{stuedi2014darpc, su2018rfp, chen2019scalerpc, monga2021flock,kalia2016fasst, kalia2019erpc}. These studies assume direct application access to network hardware and are thus
susceptible to RDMA's security weaknesses~\cite{redmark}.
\sys leverages ideas from RDMA RPC research but in a way that is
compatible with policy enforcement and observability, by doing so as a service.  Another line of work reduces the cost of marshalling, by using alternative formats~\cite{protobuf, thrift, flatbuffers, capnproto, bincode, json, msgpack, arrow} or designing hardware accelerators~\cite{karandikar2021protobuf, ourhabibi2020optimusprime, wolnikowski2021zerializer, jang2020cereal}. 
This work is largely orthogonal to our goal of removing unnecessary marshalling steps but could be applied to further
improve \sys performance.

\paragraph{Fast network stacks.}

Building efficient host network stacks is a popular research topic. 
MegaPipe~\cite{han2012megapipe}, mTCP~\cite{jeong2014mtcp}, Arrakis~\cite{peter2014arrakis},  IX~\cite{belay2014ix}, eRPC~\cite{kalia2019erpc}, and Demikernel~\cite{zhang2021demikernel} advocate building the network stack as a user-level library,
bypassing the kernel for performance. In these systems, an application directly accesses the network interface, 
but they assume policy can be enforced by the network hardware and are thus vulnerable if the hardware has security weaknesses. \sys can interpose policy on any RPC. 
Like \sys, Snap~\cite{marty2019snap} and TAS~\cite{kaufmann2019tas} implement the network stack as a service, but they stop
at layer 4 (TCP and UDP) rather than layer 7 (RPC). Application RPC stubs must marshal data into shared memory queues to use Snap
or TAS. Flexible policy engines are a key feature of Snap, but because Snap operates at layer 4, it can only apply layer 7 policies
by unmarshalling and re-marshalling RPC data.  A fast network stack like \sys can also be implemented directly in the kernel.  LITE~\cite{tsai2017lite} implements RDMA operations as system calls inside the kernel to improve manageability, and Shenango~\cite{ousterhout2019shenango} interposes a specialized kernel packet scheduler for network messages.

\paragraph{Fast network proxies.}
There is a long line of work on improving the performance of network proxies~\cite{kohler2000click,pfaff2015openvswitch,snabb,palkar2015e2,jackson2016softflow,panda2016netbricks,jamshed2017mos,katsikas2018metron,martins2014clickos,li2016clicknp,kim2015nba,miao2015silkroad,pontarelli2019flowblaze,zhang2020gallium}.
Much of this work considers the general case of a standalone proxy.  Our work differs in two ways.
First, our proposed technique is only for RPC traffic rather than generalized TCP traffic. Second, we co-design the
application library stub and proxy, and thus, both must be co-located on the same machine for our shared memory queues to function.
In today's sidecar proxies (our baseline), this assumption holds, but it does not hold for generalized network proxies.

\paragraph{Live upgrades of system software.}
Being able to update system software without disrupting or restarting applications
is key to achieving end-to-end high availability.  
Snap~\cite{marty2019snap} provides live upgrade of the network stack running as a proxy;
Bento~\cite{miller2021bento} provides similar functionality for kernel-resident file systems.
Relative to these systems, \sys upgrades are more fine-grained.  For example,
Snap targets a maximum outage during upgrades of 200 milliseconds, 
by spawning another instance of itself and moving all connections to the new process.
By contrast, our goal is near instantaneous changes and upgrades to RPC protocol definitions,
policy engines, and marshalling code. We accomplish this by keeping the control plane intact
and performing updates by loading and unloading dynamic libraries.  eBPF is a Linux
kernel extensibility mechanism that supports dynamic updates~\cite{ebpf}; unlike
eBPF, \sys can dynamically change the execution graph of policy engines as well as the individual engines
themselves.
\section{Conclusion} \label{sec:conclusion}
Remote procedure call has become the de facto abstraction for building distributed applications in datacenters. The increasing demand for manageability makes today's RPC libraries inadequate. Inserting a sidecar proxy into the network datapath allows for manageability but slows down RPC substantially due to redundant marshalling and unmarshalling. We present \sys, a novel architecture to implement RPC as a managed service to achieve both high performance and manageability. \sys eliminates the redundant marshalling overhead by applying policy to RPC data before marshalling and only copying data when necessary for security. This new architecture enables live upgrade of RPC processing logic and new RPC scheduling and transport methods to improve performance. We have performed extensive evaluations through a set of micro-benchmarks and two real applications to demonstrate that \sys enables a unique combination of high performance, policy flexibility, security, and application-level availability.
Our source code is available at \url{https://github.com/phoenix-dataplane/phoenix}.

\section*{Acknowledgement}
We thank our shepherd Amy Ousterhout and other anonymous reviewers for their insightful feedback. Our work is partially supported by NSF grant CNS-2213387 and by gifts from Adobe, Amazon, IBM, and Meta.

\bibliographystyle{plain}
\bibliography{ref}

\appendix

\clearpage

\section{Appendix}
\subsection{\sys with Full gRPC-style Marshalling} \label{sec:eval-mrpc-proto}

\begin{figure}[t]
\centering
\vspace{-1mm}
\includegraphics[scale=0.27]{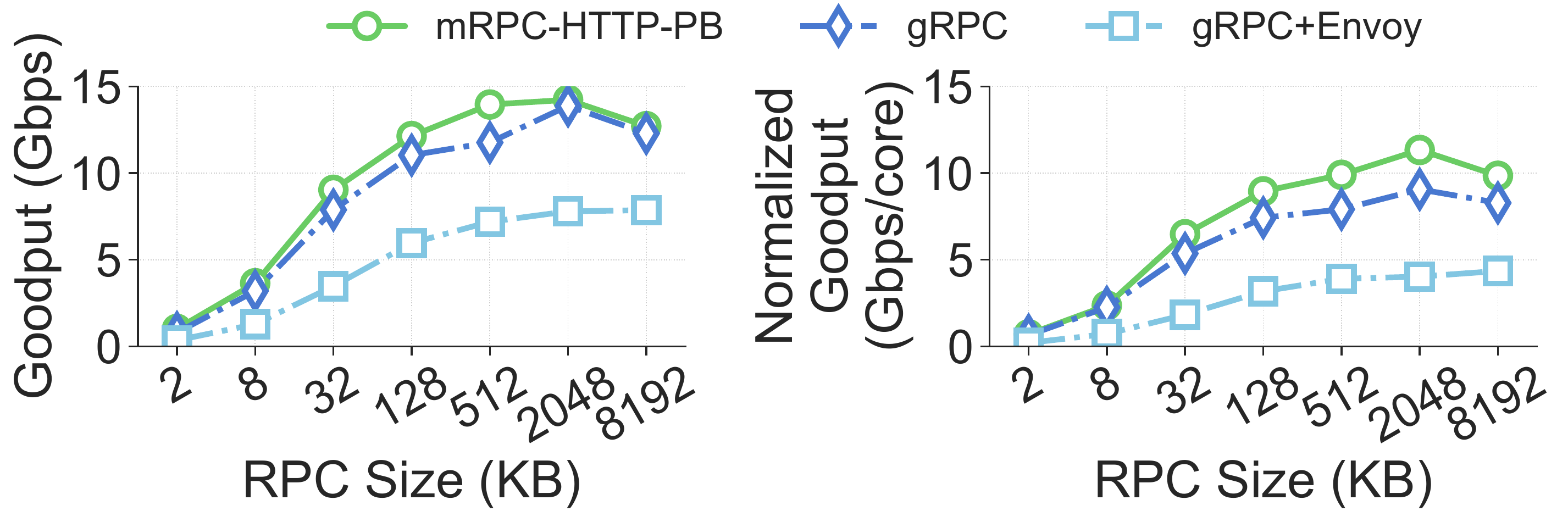}
\caption{\textbf{Microbenchmark [Large RPC bandwidth]:} Comparison of large RPC bandwidth where we use HTTP/2 and protobuf (PB) marshalling for \sys, on TCP transport. The error bars show the 95\% confidence interval, but they are too small to be visible.}
\vspace{-3mm}
\label{fig:micro-bandwidth-proto}
\end{figure}

\begin{figure}[t]
\centering
\vspace{-1mm}
\includegraphics[scale=0.28]{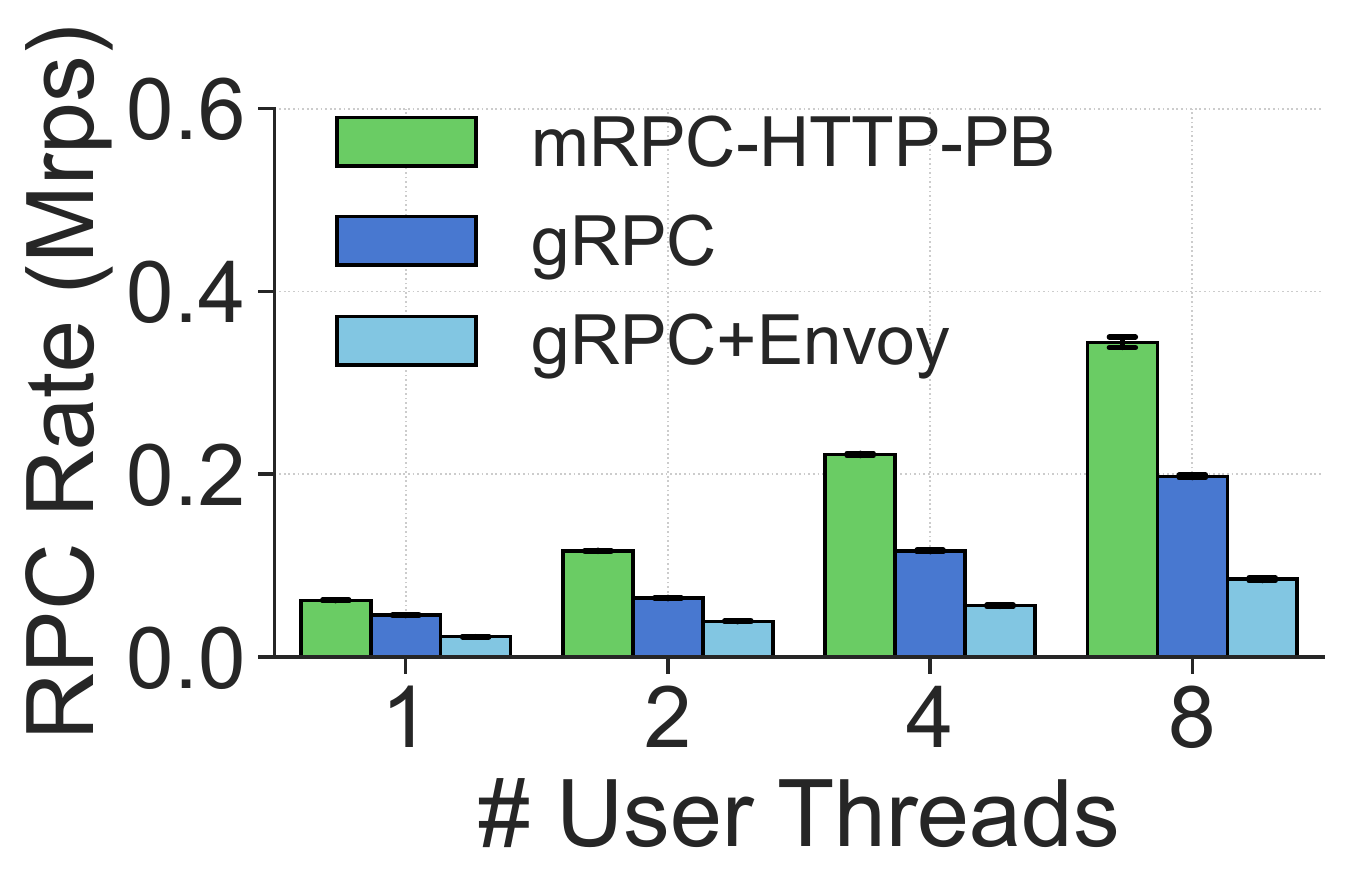}
\caption{\textbf{Microbenchmark [RPC rate and scalability]:} Comparison of small RPC rate and CPU scalability where we use HTTP/2 and protobuf (PB) marshalling for \sys, on TCP transport. The error bars show the 95\% confidence interval.}
\vspace{-3mm}
\label{fig:micro-rate-proto}
\end{figure}

As gRPC uses \texttt{protobuf}~\cite{protobuf} for encoding and HTTP/2 as the payload carrier, it has a memory copying and HTTP/2 framing cost. On the other hand, \sys is agnostic to the marshalling format. Although \sys's default marshalling is zero-copy and is generally faster than gRPC-style marshalling, our main goal of the paper is to show that we can eliminate the redundant (un)marshalling steps while enabling network policies and observability for RPC traffic. 

To isolate the performance benefits of using zero-copy marshalling and reducing the number of (un)marshalling steps, we evaluate \sys with full gRPC-style marshalling (protobuf + HTTP/2). We implement an \sys variant that applies encoding (decoding) code generated by the \texttt{protobuf} compiler and HTTP/2 framing for inter-host \sysservice communication.

We conduct the same large RPC goodput experiment in \autoref{sec:micro-bench}. The results are presented in \autoref{fig:micro-bandwidth-proto}. We find that \sys achieves performance comparable to gRPC after switching to using protobuf + HTTP/2. With full gRPC marshalling, \sys still performs 2.6$\times$ and 3.7$\times$ as fast as gRPC + Envoy in terms of goodput and goodput per core. This is because \sys reduces the number of (un)marshalling steps. The small RPC rate and scalability of \sys with gRPC marshalling is also shown in \autoref{fig:micro-rate-proto}. Since encoding small RPCs with protobuf is relatively fast, the trend to the rate and scalability is similar to~\autoref{fig:bench_rate_tcp}.

\begin{figure}
\centering
\includegraphics[width=0.8\columnwidth]{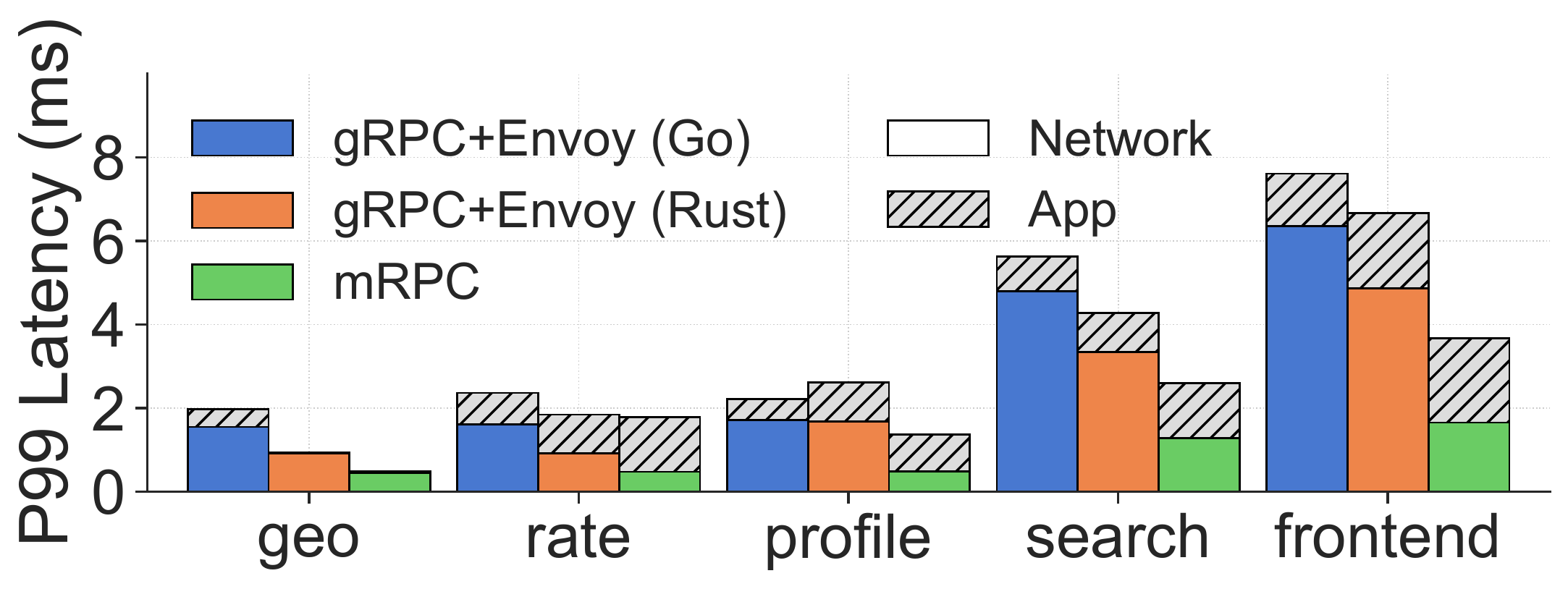}
\vspace{-2mm}
\caption{\textbf{DeathStarBench}: P99 latency of in-application processing and network processing of microservices, respectively. gRPC with Envoy and \sys are compared. A null policy is applied for \sys.} 
\vspace{-1mm}
\label{fig:deathstarbench-proxy-tail}
\end{figure}

\begin{figure}
\centering
\includegraphics[width=0.8\columnwidth]{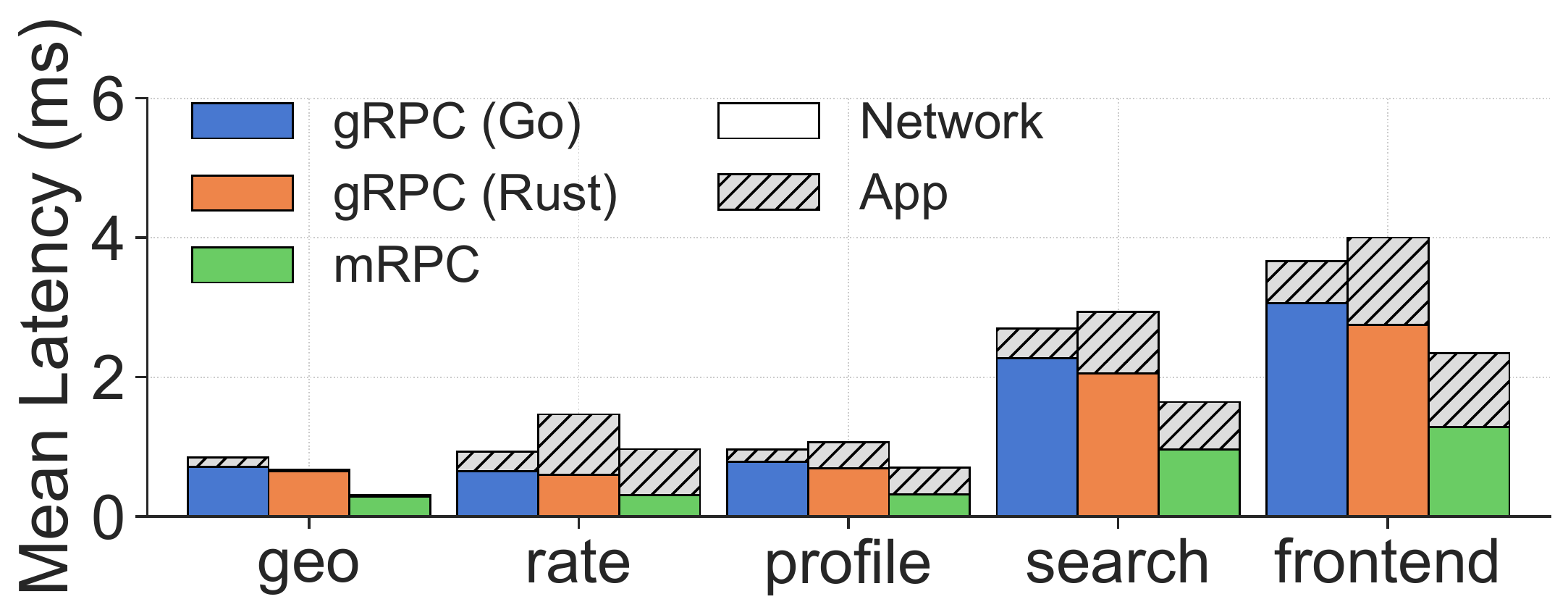}
\vspace{-2mm}
\caption{\textbf{DeathStarBench}: Mean latency of gRPC without proxy and \sys. } 
\vspace{-1mm}
\label{fig:deathstarbench-noproxy-mean}
\end{figure}

\begin{figure}
\centering
\includegraphics[width=0.8\columnwidth]{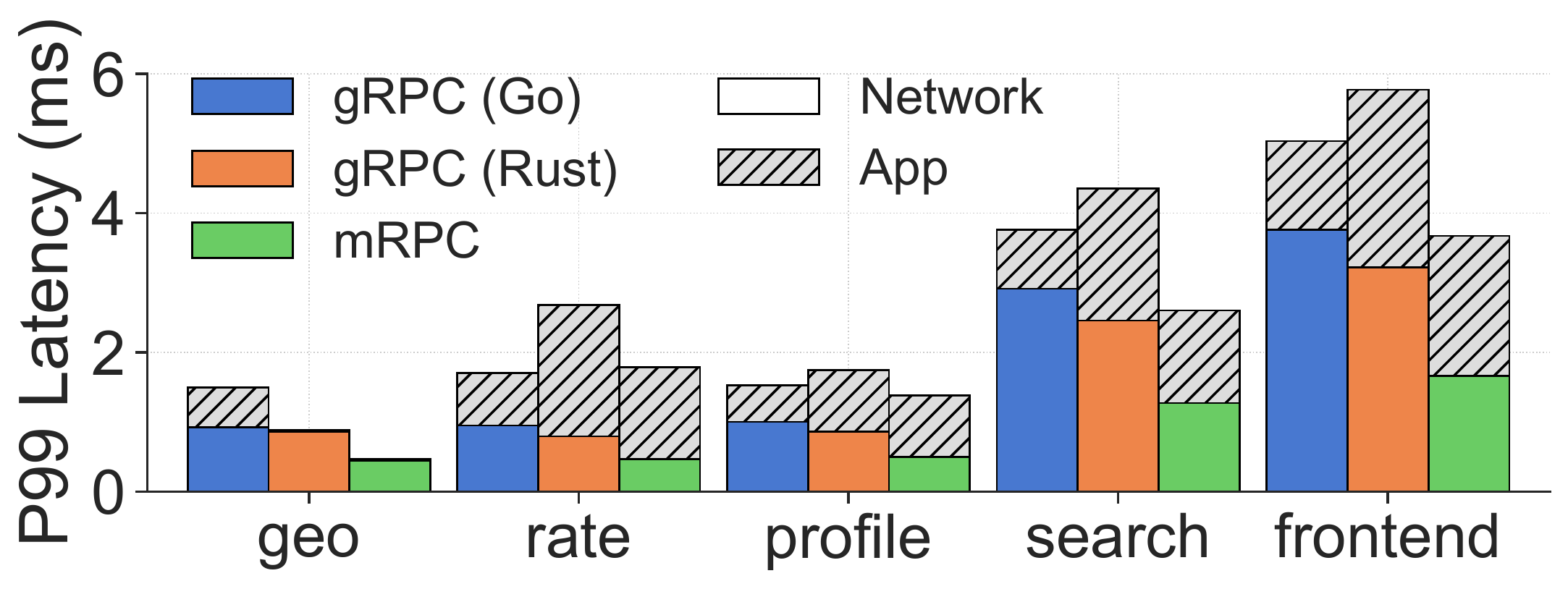}
\vspace{-2mm}
\caption{\textbf{DeathStarBench}: P99 latency of in-application processing and network processing of microservices, respectively. gRPC without proxy and \sys are compared.}
\vspace{-1mm}
\label{fig:deathstarbench-noproxy-tail}
\end{figure}

\begin{figure}
\centering
\includegraphics[width=0.8\columnwidth]{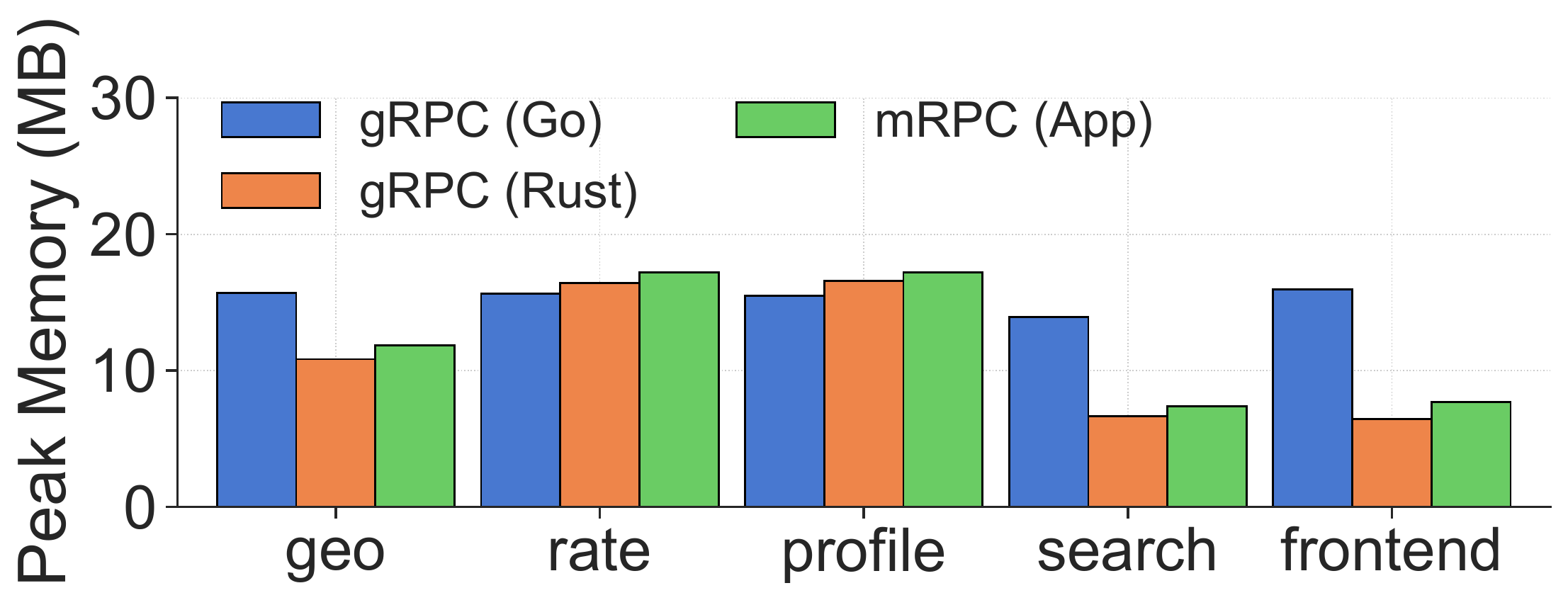}
\vspace{-2mm}
\caption{\textbf{DeathStarBench}: Peak memory usages of different services. gRPC without proxy and \sys are compared.}
\vspace{-1mm}
\label{fig:deathstarbench-noproxy-mem}
\end{figure}

\subsection{Extended Evaluation for DeathStarBench} \label{sec:real-app-appendix}

We report the P99 latency of DeathStarBench in \autoref{fig:deathstarbench-proxy-tail}, comparing gRPC with Envoy and \sys. The result is similar to the comparison of median latency in \autoref{sec:real-apps}. \sys speeds up gRPC+Envoy by 2.1$\times$ in terms of end-to-end P99 tail latency.

We also evaluate gRPC without proxy and 
\sys without any policy enforced. \autoref{fig:deathstarbench-noproxy-mean} and \autoref{fig:deathstarbench-noproxy-tail} show the results for mean latency and P99 tail latency. We observe that \sys speeds up gRPC by 1.7$\times$ and 1.6$\times$, in terms of mean latency and P99 tail latency. Communication costs are substantial in the DeathStarBench applications, and thus reducing the communication latency can improve end-to-end application performance. This is consistent with the original DeathStarBench paper's observation~\cite{gan2019deathstarbench}.

We further compare the memory usage of gRPC and \sys. The peak memory consumption of gRPC and \sys in DeathStarBench applications is illustrated in \autoref{fig:deathstarbench-noproxy-mem}. For \sys, we report the user application side memory usage, which also includes all the memory pages shared with the \sysservice. We observe that \sys does not incur 
notable memory overhead compared to gRPC. On the other hand, we find a small and constant memory footprint of \sysservice across all machines at around 9 MB.

\end{document}